\input psfig.sty
\documentclass{article}
\usepackage{graphicx,rotating,color}
\usepackage[draft]{kmn}
\usepackage{mnras_cite}

\def\spose#1{\hbox to 0pt{#1\hss}}
\def\sat{{\it RXTE}}
\def\til{$\sim$}
\def\targ{XB\thinspace1916--053}
\def\simlt{\mathrel{\spose{\lower 3pt\hbox{$\mathchar"218$}}
     \raise 2.0pt\hbox{$\mathchar"13C$}}}
\def\simgt{\mathrel{\spose{\lower 3pt\hbox{$\mathchar"218$}}
     \raise 2.0pt\hbox{$\mathchar"13E$}}}
\def\degree{\nobreak\ifmmode{^\circ}\else{$^\circ$}\fi}

\renewcommand{\sec}{\thinspace\hbox{s}}
\newcommand{\persec}{\thinspace\hbox{s$^{-1}$}}
\newcommand{\dy}{\thinspace\hbox{d}}

\newcommand{\pd}{\thinspace\hbox{d$^{-1}$}}
\newcommand{\msun}{\thinspace\hbox{$M_{\odot}$}}
\newcommand{\rsun}{\thinspace\hbox{$R_{\odot}$}}

\title[On the multi-periodicities in the X-ray dipper XB\thinspace1916--053]
{On the multi-periodicities in the X-ray dipper XB\thinspace1916--053}
\author[L. Homer et al.
]
	{L. Homer$^1$\thanks{Present address: Astronomy Department, University of Washington, Box 351580, Seattle WA
	98195-1580, USA. \hspace{2cm}email: lh@astro.ox.ac.uk.  }, P. A. Charles$^{1,3}$, P. Hakala$^{2}$, P. Muhli$^2$,  I-C. Shih$^1$, \\
\hspace{5mm}\\
{\LARGE \rm A. P. Smale$^4$ and G. Ramsay$^5$}\\
\hspace{5mm}\\
$^1$Department of Astrophysics, Nuclear Physics Laboratory, Keble Road, Oxford OX1 3RH, UK\\
$^2$Observatory and Astrophysics Laboratory, P. O. Box 14, FIN-00014, University of Helsinki, Finland\\
$^3$Department of Physics \& Astronomy, University of Southampton, 
 Southampton, SO17 1BJ, UK\\
$^4$Laboratory for High Energy Astrophysics, Goddard Space Flight Center, Greenbelt, MD 20771, USA\\
$^5$Mullard Space Science Laboratory, University College London, Holmbury St.Mary, Dorking, Surrey RH5 6NT, UK }
\date{Accepted 1999 December 31. Received 1999 January; in original
form \today}

\begin{document}

\maketitle

\begin{abstract}
Using the {\it Rossi X-ray Timing Explorer} and the Nordic Optical Telescope we have obtained the highest ever quality X-ray/white-light high-speed photometry of
XB\thinspace1916--053.  We refine the X-ray period ($P_X$) to $3000.6\pm0.2$\sec\ via a
restricted 
cycle counting approach.  Using our complete optical lightcurve, we have extended the optical period ($P_{opt}$) ephemeris by another 4
years, providing further evidence for its stability, although a slightly longer period
of $3027.555\pm0.002$\sec\ now provides a marginally better fit.  Moreover, modulations at both $P_X$ and $P_{opt}$ are present in the optical
data, with the former dominating the nightly lightcurves (i.e. a few cycles of data).  We have also attempted to determine the ``beat'' period, as seen in the repeating evolution of the X-ray dip structure, and the
variation in
primary dip phase.  We find that a quasi-period of $4.74\pm0.05$\dy\ provides the best fit to the data, even then requiring phase shifts
between cycles, with the expected 3.90\dy\
``beat'' of $P_X$ and $P_{opt}$ appearing to be less likely.
	Finally, considering the nature of each of these temporal phenomena, we outline possible models, which could explain all of the
observed behaviour of this enigmatic source, focusing on which of $P_X$ or $P_{opt}$ is the binary period.
\end{abstract}


\section{INTRODUCTION}
The low-mass X-ray binary (LMXB) \targ\ was the first of the 10 dipping sources to be discovered.  The recurrence period of the dip \til 3000~s
\cite{walt82,whit82a} also provided one of the first binary period determinations for a LMXB, and places it as the shortest period dipper.  It also exhibits type-I
X-ray bursts confirming that the compact object is a neutron star.  A low-mass (\til0.1\msun)  donor is then required to fit into the system, which is most
likely evolved, partially degenerate and hydrogen depleted \cite{nels86}.  The lack of eclipses by this companion, for a representative radius \til0.1\rsun, yields an
inclination in the range 60\degree\ to 79\degree\ \cite{smal88}.

The generally accepted model for the origin of X-ray dips, suggests that they are due to the obscuration of the central source by vertically
extended material at or close to the outer edge of the accretion disc \cite{parm88,armi98}.  In recent years Church and co-workers have studied a selection of
dippers, utilising the superior spectral capabilities of {\it ASCA} and {\it BeppoSAX}, to probe in detail the spectral evolution through the
dips \cite{chur93,chur95}. They find that a partial covering absorption model reproduces
the observed spectra well, where the fractional contribution from each of two components (point-like blackbody
and more spatially extended cut-off power law) changes independently.  For \targ\ a physical model is implied consisting of the point-like blackbody emission region (presumably
the neutron star surface and/or adjacent boundary layer), plus the cut-off power law component, originating in an extended accretion disc
corona (ADC, responsible for Comptonization, \pcite{chur97}).   In the non-dip spectra the blackbody alone contributes  \til30\% to the total
flux \cite{chur97,chur98}.  During dips the blackbody component
is very rapidly and totally obscured, whilst the power-law is more gradually attenuated, as the partial-covering fraction increases (even to
complete coverage in the deepest dips seen).
From measurement of the ingress/egress times a size for the radial extent of the ADC is found of 20\% of the tidal disc radius.

   \begin{figure}[!htb]
      \centering
      \includegraphics[scale=0.41,angle=0]{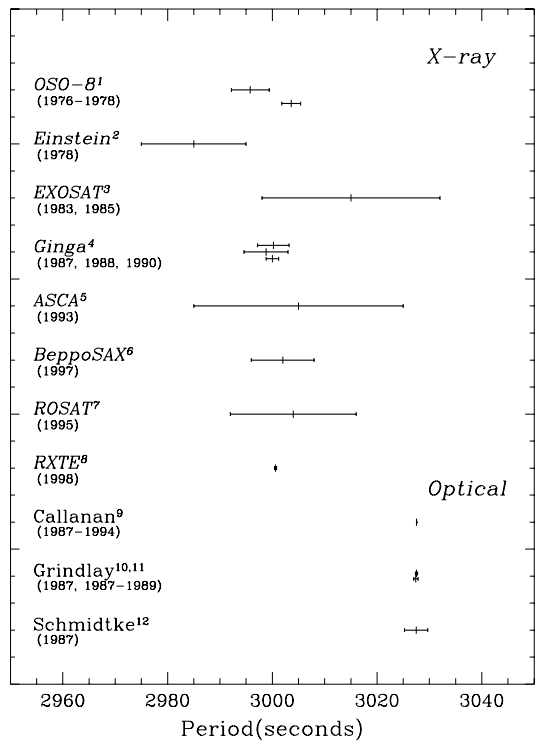}
      \caption{Determinations of the ``orbital'' period of \targ.  The X-ray period
      ($\sim$3000s) measurements from a long-line of satellite missions at the top and the
      three previous optical results at the bottom {(\small 1. White \& Holt 1982, 2. Walter {\em et al.} 1982, 3. Smale {\em et al.} 1989,
      4. Yoshida {\em et al.} 1995, 5. Church {\em et al.} 1997, 6. Church {\em et al.} 1998, 7. Morley {\em et al.} 1999, 8. this paper, 9. Callanan, Grindlay \& Cool 1995,
      10. Grindlay {\em et al.} 1988, 11. Grindlay 1989, 12. Schmidtke 1988).} \label{fig:Pdets}}
    \end{figure}

\subsection{Multi-periodicities of \targ}
Following their discovery of an optical counterpart to \targ\ \scite{grin88} searched for and found an optical modulation close to
the 3000\sec\ dip recurrence period.  Subsequently, this source has been the object
of many timing studies.  It was soon apparent that the periods derived from X-ray and optical observations were inconsistent, differing by \til
1\%.  The results of these determinations are summarised in Fig.\ref{fig:Pdets}.  For the X-ray dips, the earliest method adopted (by \pcite{walt82,whit82a}), was to consider the timings of a series of (near-consecutive) dips from a single satellite pointing,
and calculate the best fit recurrence time.  Unfortunately, the short observing baseline leads to large uncertainties in the period. More recently, the approach generally taken has been to
epoch-fold the lightcurves on a grid of trial periods, and compute chi-squares to identify the period yielding the most distinct modulation. However,
given the notable variation in the morphology of the dip structure over time (see e.g. \pcite{smal88,smal92b,yosh95,chur97}), the precision of
this method is also
uncertain, and the large errors quoted may still be underestimates.  In their optical studies, \scite{schm88} and \scite{grin88} used a
straightforward Fourier transform of their long (and irregularly) sampled lightcurves, whilst \scite{call95} refined this by way of cycle-counting using the minima in the folded optical lightcurves to define the ephemerides.  

Not only are two close periodicities present, there is evidence for two longer periodic phenomena.  In fact, it is assumed that physically one of
the two $\sim3000$s periods arises from binary motion, whilst the other is due to a beat with another longer period effect, which must be close to 4
days.  Indeed changes in the lightcurve shapes in both bands and in particular the phases of dip features has
been seen on this timescale \cite{grin89,smal89,yosh95}.  Note, for brevity, we shall henceforth refer to any \til4\dy\ period as the ``beat''
period of $P_X$ and $P_{opt}$, whilst reserving the notation $P_{beat}$ for the quantity calculated as $(P_X^{-1} -P_{opt}^{-1})^{-1}$, and using $P'_{beat}$ for any other period close to this, i.e.
not assuming that they are the same.  Lastly, an analysis of {\it Vela 5B} data by \scite{prie83} yielded a possible longer-term variation in X-ray flux
of 199 days.  But with a false alarm probability of 10-20\% \cite{smal92}, this detection was by no means firm.

A number of models have been proposed to explain the multi-periodicities, as we shall discuss in section~\ref{sect:disc}, but no clear consensus has been achieved given the
limited constraints imposed by the observations to date.  It was clear that further complementary X-ray and optical observations were vital in order to solve the mystery of the multiple periodicities and
constrain the range of proposed models to explain them.  Hence, we obtained time with both the {\it Rossi X-ray Timing Explorer (RXTE)} and
the Nordic Optical Telescope (NOT) to undertake a 
simultaneous X-ray/optical campaign.  It is the results of these observations that we present here.  In section~\ref{sect:ObsDR} we summarise the observations and data
reduction performed, moving on to the temporal analysis in \S~\ref{sect:Tanal1} (X-ray period), \S~\ref{sect:Tanal2} (``beat'' period),
\S~\ref{sect:Tanal3} (longer-term variations) and \S~\ref{sect:Tanal4} (optical periods).  We discuss our results in section~\ref{sect:disc}.

\section{Observations and data reduction}
\label{sect:ObsDR}

\begin{table}[htb!]
\caption{X-ray and optical observations of \targ\ \label{tab:obslog}}
\begin{tabular}{l l c c} 
Instrument/ &Date (UT) & \multicolumn{2}{c}{Time span}\\
Band						& 				 &(h)&(binary\textcolor{white}{)} \\
							&				&		&\textcolor{white}{(}orbits)\\
{\em RXTE}/PCA		&   23.06.98 &  0.64  & 0.76\\
X-ray (2--60 keV)  &   24.06.98 &  0.60  & 0.72 \\
				 		&	 25.06.98 &  0.60  &  0.72 \\
				 		&	 26.06.98 &  0.59  &  0.70 \\
				 		&	 27.06.98 &  0.27  &  0.33 \\
				 		&	 17.07.98 &  5.89  &  7.07 \\
				 		&	 18.07.98 &  8.95  & 10.74 \\
				 		&	 19.07.98 & 11.31  & 13.57 \\
				 		&	 20.07.98 &  9.65  & 11.58 \\
				 		&	 21.07.98 &  5.72  &  6.87 \\
				 		&	 23a.07.98&  4.29 &   5.15 \\
				 		&	 23b.07.98&  6.58  &  7.89 \\
				 		&	 24.07.98 &  4.04  &  4.85 \\
				 		&	 25.07.98 &  4.05  &  4.86 \\
				 		&	 26.07.98 &  3.82  &  4.59 \\
				 		&	 01.08.98 &  3.61  &  4.33 \\
				 		&	 10.08.98 &  2.85  &  3.42 \\
				 		&	 14.09.98 &  1.28  &  1.54 \\
				 		&	 16a.09.98&  0.23  &  0.28	\\
				 		&	 16b.09.98&  0.94  &  1.13	\\
				&&&\\
Nordic Optical & 	   25.06.98 & 2.29 & 2.75 \\
Telescope/CCD		&	25/26.06.98 & 1.94 & 2.33 \\
 white light		&	26/27.06.98 & 1.94 & 2.33 \\
				 		&	27/28.06.98 & 1.94 &2.33 \\
				 		&	29.06.98 & 2.11 & 2.53  \\
				&&&\\
{\em RXTE}/ASM 		&24.02.96 & 1355d &-  \\
X-ray (2--10 keV)& --09.11.99&\\
\end{tabular}
\end{table}  
 
\begin{figure*}[!htb]
      \centering
      \includegraphics[scale=0.6,angle=-90]{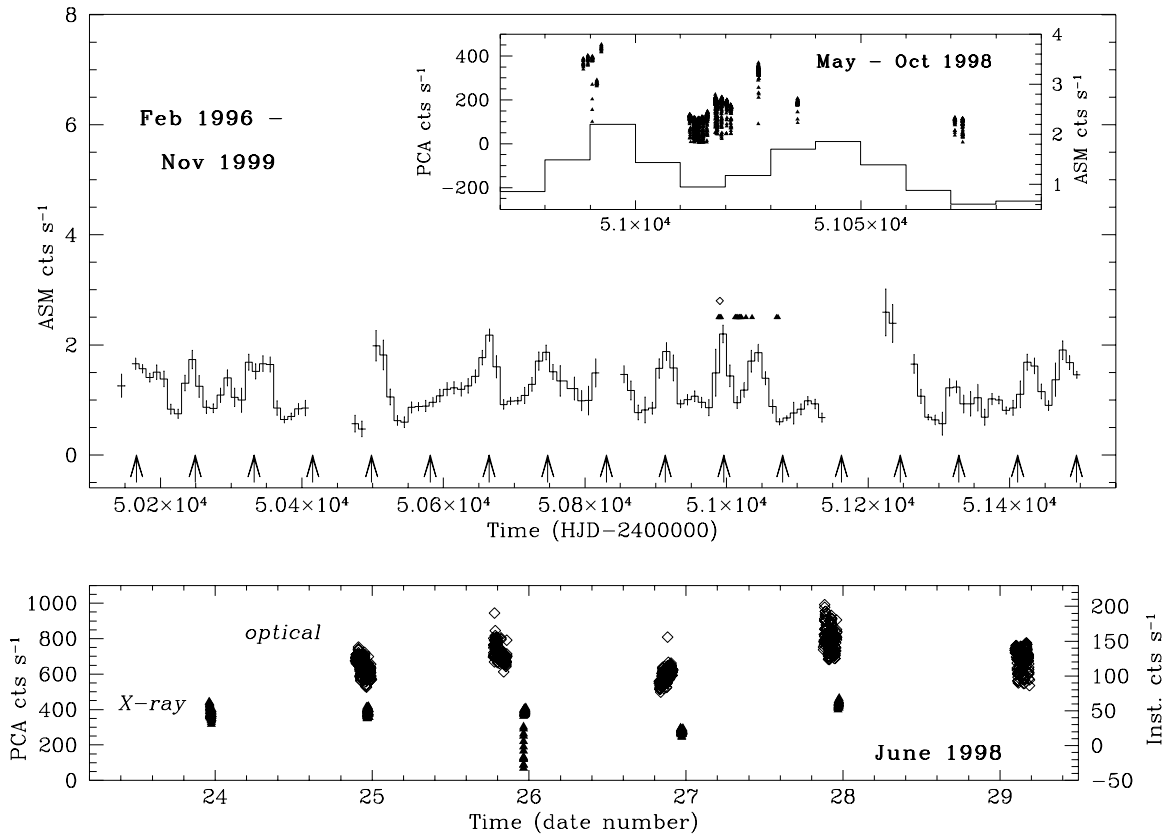}
      \caption{Temporal overview of the various datasets used in this paper.  Upper panel: the 10\dy-averaged lightcurve from the \sat/ASM with times of the pointed observations marked with symbols: solid triangles-- X-ray (\sat/PCA), open
diamond-- optical (NOT).  The vertical arrows
indicate the predicted times of maximum X-ray flux, according to the 83 day (quasi?) periodic modulation found in the long-term ASM X-ray
lightcurve.  Inset: detail of the X-ray lightcurves from all the PCA observations (points), together with the ASM monitoring (histogram).
Lower panel: detail of the simultaneous observations in June 1998 with the optical lightcurve shown uppermost (open diamonds) and the X-ray
(PCA) below (solid triangles).\label{fig:obslog}}
    \end{figure*}

\targ\ and its optical counterpart were observed partly simultaneously using \sat\ and NOT, La Palma in 1998 June.  Unfortunately, the coverage
available by the satellite was very restricted, and further extensive observations were obtained in the period 1998 July
17-September 16.  In addition we have considered the long-term flux behaviour as measured by the All-Sky Monitor (ASM) on-board \sat. The log
of all these observations is given in table~\ref{tab:obslog}, whilst their relative distribution
and overlap is shown in Fig.\ref{fig:obslog} (together with a plot of the the long-term ASM lightcurve).

\subsection{\textit{RXTE}/PCA}
For our short-term variability analysis we made use only of data taken with the Proportional Counter Array (PCA)
instrument \cite{brad93} on {\it RXTE}.  This consists of five nearly
identical Proportional Counter Units (PCUs) sensitive to X-rays with
energy range
between 2--60 keV and a total effective area of $\sim$ 6500\,cm$^2$. The combination of large area and an energy
range covering the peak of the X-ray spectrum of \targ, means it provides good signal-to-noise lightcurves.  Since
the source is an X-ray burster the
PCA data were obtained in two triggered high time-resolution burst modes, so that X-ray bursts could also be studied
in detail (the subject of a future paper), as well as in both the standard modes.  However, for our study of the periodic temporal phenomena presented
here, we excluded the bursts.  The `standard2', 16s time resolution data were extracted using {\small XSELECT}, and
then background-subtracted using the Epoch 3 bright source model.  Given the average count-rate in the PCA of 100--300 cts~s$^{-1}$, this correction should be
perfectly adequate.  Lastly, the timing information was transformed into HJD, to match up with that of the optical data.

\subsection{NOT}

The optical counterpart to \targ\ was observed using the Nordic Optical Telescope (NOT), La Palma,
during 5 consecutive nights starting 25 June 1998. Each night 
the observations covered at least two orbital periods (i.e. 2 hours
minimum per night). 
	The observations were obtained using ALFOSC, which is a
multi-purpose medium/low resolution spectrograph and imager. ALFOSC was
equipped with a 2kx2k Loral-Lesser thinned AR-coated CCD detector, which
was operated in a sub-windowed mode enabling a readout time of the
order of 3\sec. This, coupled with exposure times of 10\sec\ 
then yields a true time resolution of $\sim$ 13\sec. All the
observations were made in white light for maximum signal to
noise and best possible time resolution. Note the Loral-Lesser
chip has an almost flat high (75-85\% ) QE over most of the optical range. All the images were then bias subtracted and flatfielded in
the usual manner.
	As the target is located just $\sim$2\arcsec away from a 2 mag brighter
star good seeing is vital for the photometry.
During our run the seeing was generally better than 1.0", which allowed us to
use the IDL implementation of the DAOPHOT routines to reliably extract the light 
curves of the target and the 2 comparison stars within the same frame.
Differential photometry was then performed to obtain the final lightcurves.

\subsection{\textit{RXTE}/ASM}
Unsurpassed long-term monitoring of the flux from all the bright X-ray sources is available from the All-Sky Monitor (ASM) of \sat\ .  This
consists of three wide-angle shadow cameras equipped with position sensitive Xenon proportional counters with a total collecting area of 90 cm$^2$.  They work in the
2-10 keV energy range, providing roughly a timing resolution of 90 minutes between dwells (i.e. 80\% of the sky is scanned every 90 minutes on average), and are sensitive
to sources as faint as 30 mCrab. 

We extracted the complete lightcurve of \targ\ (spanning 1355 days) from the HEASARC database (the definitive flux values for each dwell), which
contains both total flux and that in three energy bands.  The average
count rate is only about 1.2\persec (16 mCrab), hence rebinning is required to obtain adequate signal-to-noise.  We derived both 1 day and 10
day averages, using the same screening
as that performed by the MIT ASM team, but also a requirement that there were at least 10 (100) dwells per bin.  This latter criterion
ensured all data points had approximately the same estimated uncertainty, and generally excluded poor data.  However, we also found it necessary to exclude
(by hand)
the block of data from December 1998 -January 1999, when the viewing of \targ\ led to particularly poor data.  The effect of restricted viewing
was also evident in a
number of other galactic centre X-ray binaries; each exhibited significant and variable increased flux levels during these months.

\section{Refining the X-ray period}
\label{sect:Tanal1}

Given the well-documented morphology changes seen in the X-ray lightcurves of \targ\, we firstly decided to examine the dip behaviour in our
      own \sat/PCA data.  However, due to the short time ($\simlt2700$~s) on source per satellite orbit we had to use the combined lightcurve from a few such
      segments and phase fold, to provide full phase coverage of the dip period.   \begin{figure*}[!htb]
      \centering
      \includegraphics[scale=0.6,angle=-90]{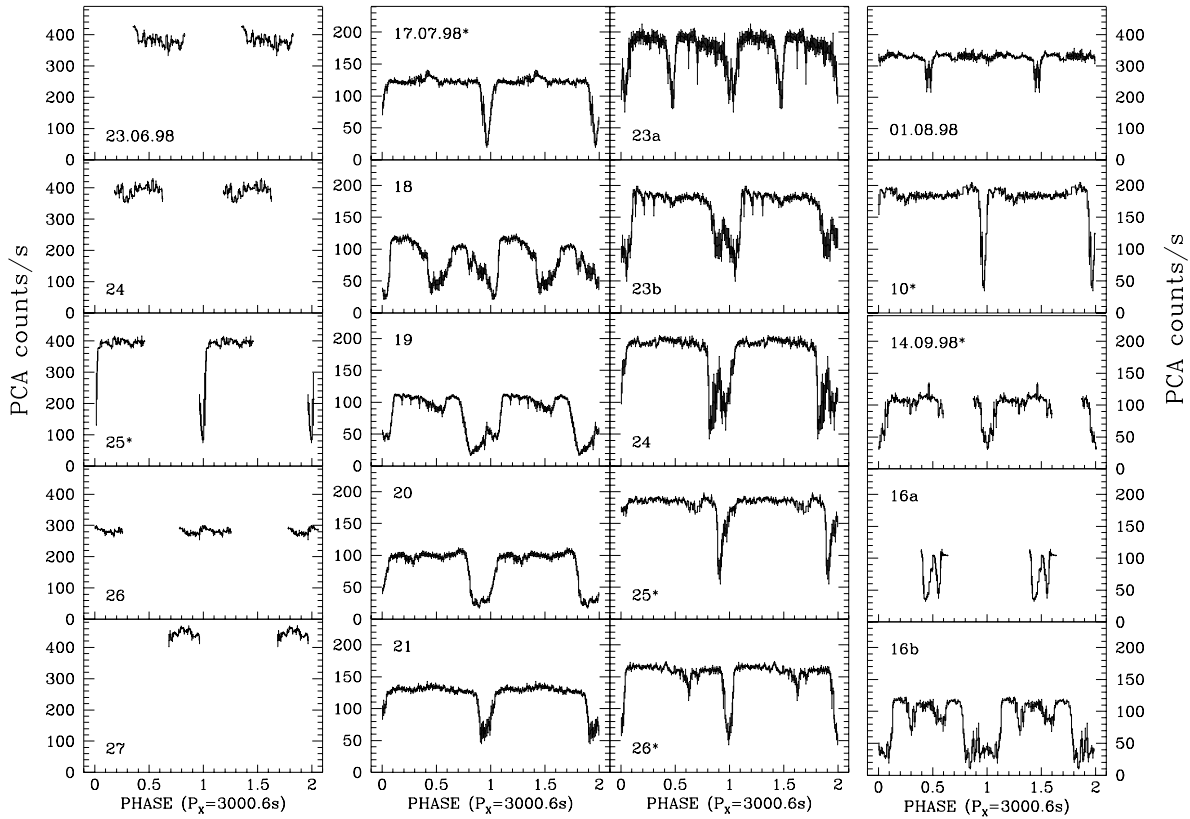}
      \caption{ The PCA X-ray lightcurves folded on the refined $P_X=3000.6$s, with 40 phase bins.  Each panel shows the data from a single pointing
      (which occurred at most daily).  In two cases, the pointing has been first split into two subsets, as we noted dramatic changes in
      morphology during that time.  For clarity two full cycles are shown, and different scalings of the count rate axes have been employed.  Note the clearly evolving morphology, which returns to
      approximately the same state every 4--5 days.  The lightcurves marked by * were used to define ephemerides for cycle counting (see
      text). \label{fig:xfolds}}
    \end{figure*}
    Hence, we
first assumed a representative X-ray period of $3000.0$~s (see Fig.~\ref{fig:Pdets}), and folded the data from each day's observations on this period.  This procedure
produced in most cases a good signal-to-noise average lightcurve (after phase binning), with complete phase coverage.  The final set
 of resulting
folded lightcurves (following the period refinement) are presented in Fig.\ref{fig:xfolds}.

\begin{figure*}[!htb]
      \centering
      \resizebox*{.8\textwidth}{0.35\textheight}{\rotatebox{-90}{\includegraphics{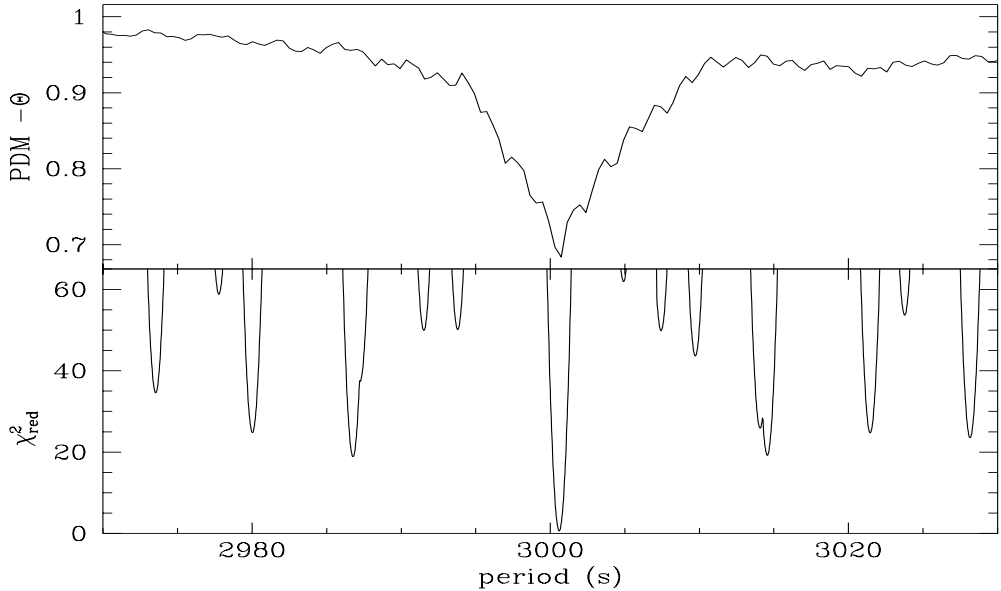}}}
      \caption{Upper panel: Phase Dispersion minimisation periodogram.  Lower panel: results of the cycle counting method -- reduced $\chi^2$
for the observed ({\it O}) - computed ({\it C}) times of dip minimum (for
narrow dip datasets), for a corresponding range of periods.  Within the range of possible periods consistent with the PDM result, there is one
minimum from the cycle counting approach, which has $\chi^2_{\nu}$\til1 far smaller than any others. \label{fig:xpdm}}
    \end{figure*}

The striking cyclical variation in morphology is immediately apparent.  It is clear given the highly non-sinusoidal nature of the modulation
that a Fourier-based period search is not appropriate.  Furthermore, even the use of the phase dispersion minimisation (PDM, \pcite{stell78})
periodogram may
not be ideal.  However, to make progress we did perform a PDM analysis and the results are shown in Fig.\ref{fig:xpdm}. This yields $P_X=3001.0\pm3.0 $~s, where the
precision is limited by the large morphological changes and the scatter they produce.

\begin{figure*}[!htb]
      \centering
		\vspace{-5cm}
      \includegraphics[scale=0.6,angle=-90]{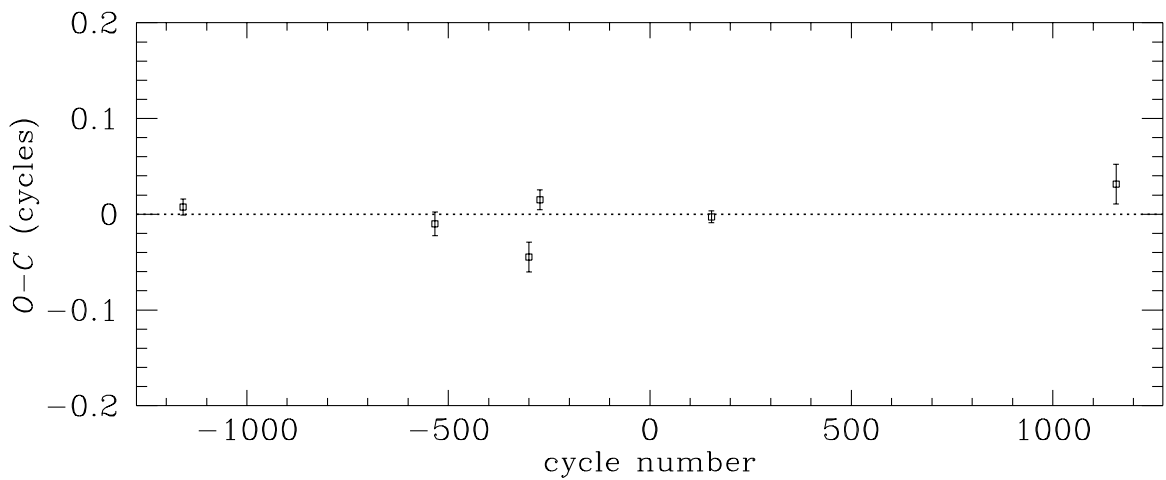}
      \caption{For the best fit period of 3000.6 s, the observed ({\it O}) - computed ({\it C}) times of dip minimum in out \sat/PCA data (for narrow dips only) are shown, in terms of
      cycles.  \label{fig:ocplot}}
    \end{figure*}

To refine this result, we attempted for the first time a cycle-counting analysis of the dip timings. Firstly, we took the period found using the PDM, and then folded and phase-binned each day's data on this, with an arbitrary phase zero ($T_0$), to derive an average, good signal-to-noise profile for the dip. Even by inspection it is clear that these phases vary
systematically with the periodic overall shape changes.  Hence, we decided to derive ephemerides for the principal dip only for folds which
appear to be at least at roughly the same phase in the ``beat'' cycle (i.e. whenever we have a single narrow dip morphology), as also
indicated (by $\ast$) in Fig.~\ref{fig:xfolds}.  Next, a Gaussian profile was fitted to
these dips to estimate their centre phase ($\phi_{dip}$), with a generous error given by the FWHM of this fit (rather than the statistical error on the centre
value).  Thence, the time for the dip minimum is given by $T_{dip,n}=T'_{0,n}+\phi_{dip}\times P_{dip}$, where
$T'_{0,n}=T_0+int[(T_{midpt}-T_0)/P_{dip}]\times P_{dip}$, in order to give a time close to the mid-point time ($T_{midpt}$) of the $n$th dataset.
We then have a set of 6 ephemerides spanning the full 3 months, which should all agree with a single recurrence period.  For each cycle and
trial period, a time of minimum is predicted ({\it C}) which is then compared to the observed time ({\it O}), and a simple $\chi^2$
minimisation for the {\it O-C} values
versus trial period then gives the best fit period of $3000.6\pm0.2$~s.  The plot of  $\chi^2$ versus $P_{trial}$ (Fig.~\ref{fig:xpdm}) does show that the data can be fit by
a number of periodicities, but only this result is consistent with the foregoing PDM analysis (not to mention the published results, see Fig.\ref{fig:Pdets}).  The O-C
plot for the best fit period is shown in Fig.\ref{fig:ocplot}, which demonstrates the very small scatter of
the dip phase for this period, over the 3 month time span! 

\section{Investigating the ``beat'' period}
\label{sect:Tanal2}
We have previously noted the evolving X-ray lightcurve morphologies, whose timescale of repetition is approximately that of the ``beat''
period.  An estimate of such a period is important as we believe the changes to be directly related to a precessing disc or whatever provides the
longer period clock in \targ.  Clearly, the energetics and extent of the disc/accretion stream interaction region (which provides the obscuring
material for the dips) is varying, even significantly within a single visit (June 23 and September 16). In principle detailed modelling of
these 
lightcurves can tell us much about the time-varying geometrical structure of the disc and how the stream impacts it.  However, in this paper we
will simply concentrate on the basic timing aspects and draw qualitative conclusions (see \S~\ref{sect:disc}).

\begin{figure*}[!htb]
      \centering
\resizebox*{0.75\textwidth}{0.75\textheight}{\rotatebox{0}{\includegraphics{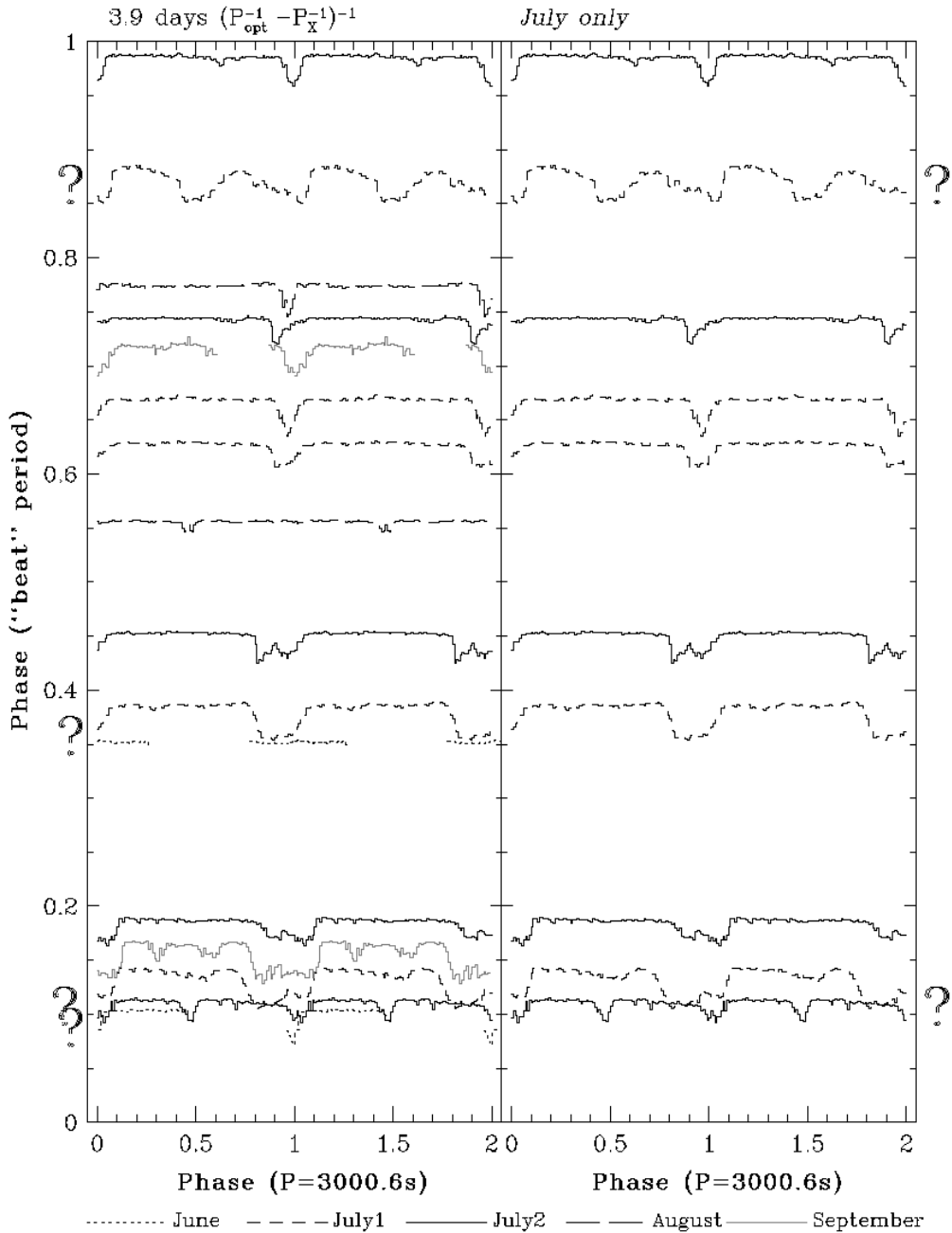}}}
      \caption{Each of the X-ray datasets folded on $P_X$ and normalised by dividing by the non-dip flux level, and then plotted as a
    function of phase in the 3.90\dy\
    ``beat'' cycle. The folds corresponding to different cycles have been marked by different line styles (see key at bottom).  The `?'
    indicate cases where the phasing is inconsistent with the expected systematic morphology evolution over a ``beat'' cycle. Right: in particular, we
    note the discrepancy between the consecutive July cycles, as replotted here for clarity.  In one cycle, there are narrow
    dip morphologies at $\phi_{beat}=0.75$ and 0.98, whilst in the next cycle the same region of phase is occupied by an example of a more complex broad, primary +
    secondary dip shape (at $\phi_{beat}=0.88$).  There is also a departure from systematic evolution at around $\phi_{beat}=0.1$. \label{fig:trailfds3.9}}
    \end{figure*}

With our significantly improved measurement of $P_X$ we may now calculate $P_{beat}=(P_X^{-1} -P_{opt}^{-1})^{-1}$ to much higher precision.
Adopting the published value of $P_{opt}=3027.551\pm0.004$\sec\ yields $P_{beat}=3.90\pm0.03$.  But is this period consistent with the repeating
changes in lightcurve morphology?  In figure~\ref{fig:trailfds3.9}, we have plotted all of the folded X-ray lightcurves, vertically offset according to their phase assuming a 3.90\dy\ period.  There are at least two cases where from
one ``beat'' cycle to the next, totally different dip morphologies would appear at the same ``beat'' phase, and hence we exclude this
period.  

\begin{figure*}[!htb]
      \centering
      \includegraphics[scale=0.6,angle=-90]{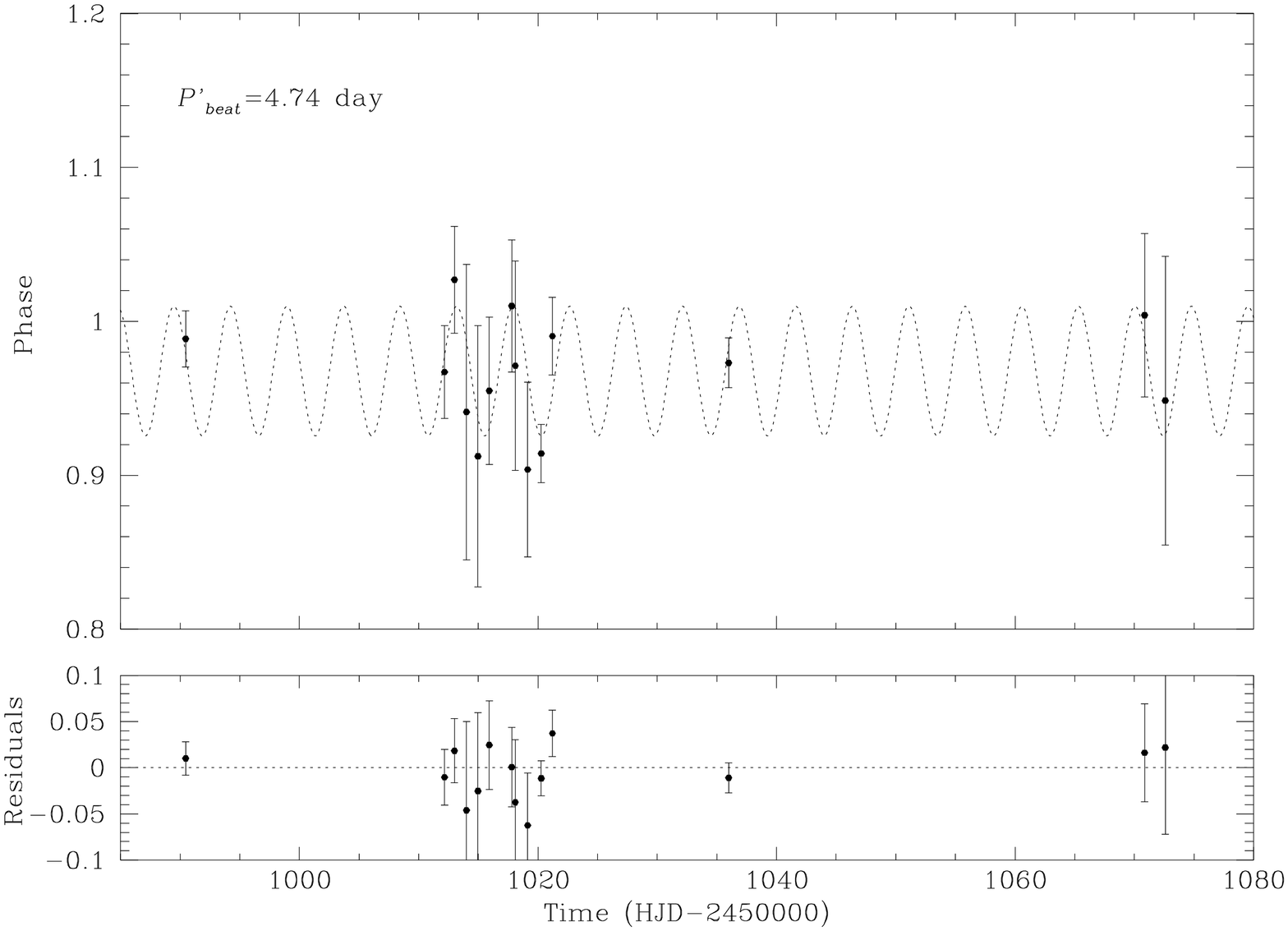}
      \caption{Considering {\em all} the \sat/PCA X-ray data which shows a primary dip, the phase of dip minimum is plotted against mid-time of
observation.  The periodic wandering in phase is most apparent. The candidate $P_{beat}'$=4.74\dy\ is shown fitted to the data, along
      with the residuals.  This fit yields an ephemeris of $T_0({\small \rm HJD})=2451030.93\pm0.27 + n*(4.74\pm0.05)$. \label{fig:ocplotPS}}
    \end{figure*}

We will now attempt to determine a ``beat'' period, which is in agreement with the  morphology changes.  However, considering when the narrow
dips or any other identifiable forms repeat may only crudely constrain the period to between about 3.5 and 5
days.  We need another more quantitative method to indicate candidate periods.  In principle there are a number of ways to do this, namely: area
of dip (i.e. below non-dip level), length of a dip, depth of dip, and phase of deepest dip.  Having tried each, and found no clear indication of the known $\sim$4 day cycle, apart from in the case of the
phase measure, we settled on this last method.  The phase and error were determined as for the narrowest dips previously.  Of course in some
cases the dip morphology is very complex, and we had to subjectively decide which part to fit with the Gaussian, although the generous FWHM error estimates will largely take this source of uncertainty into account. In the case of an asymmetric dip, the fit was only done on the portion which
was roughly Gaussian in shape.  The resultant dip phases as a function of time (the midpoint of each observing block) are plotted in
Fig.\ref{fig:ocplotPS}. 

Owing to the restricted sampling, a number of sinusoids with different periods can be fitted to the data.  We used $\chi^2$ fitting to take the dip
phase uncertainties into account, finding local minima in the expected range at 5.08\dy ($\chi^2_\nu$=0.97, 10 d.o.f.), 4.74 (0.59), 4.39 (0.01),
4.10 (0.98), and 3.84 (0.88)d.   We do note that the shortest period fit is $3.84\pm0.06\dy$ which
is therefore within 1$\sigma$ of $P_{beat}$=3.90\dy. 
\begin{figure*}[!htb]
      \centering
\resizebox*{0.75\textwidth}{0.75\textheight}{\rotatebox{0}{\includegraphics{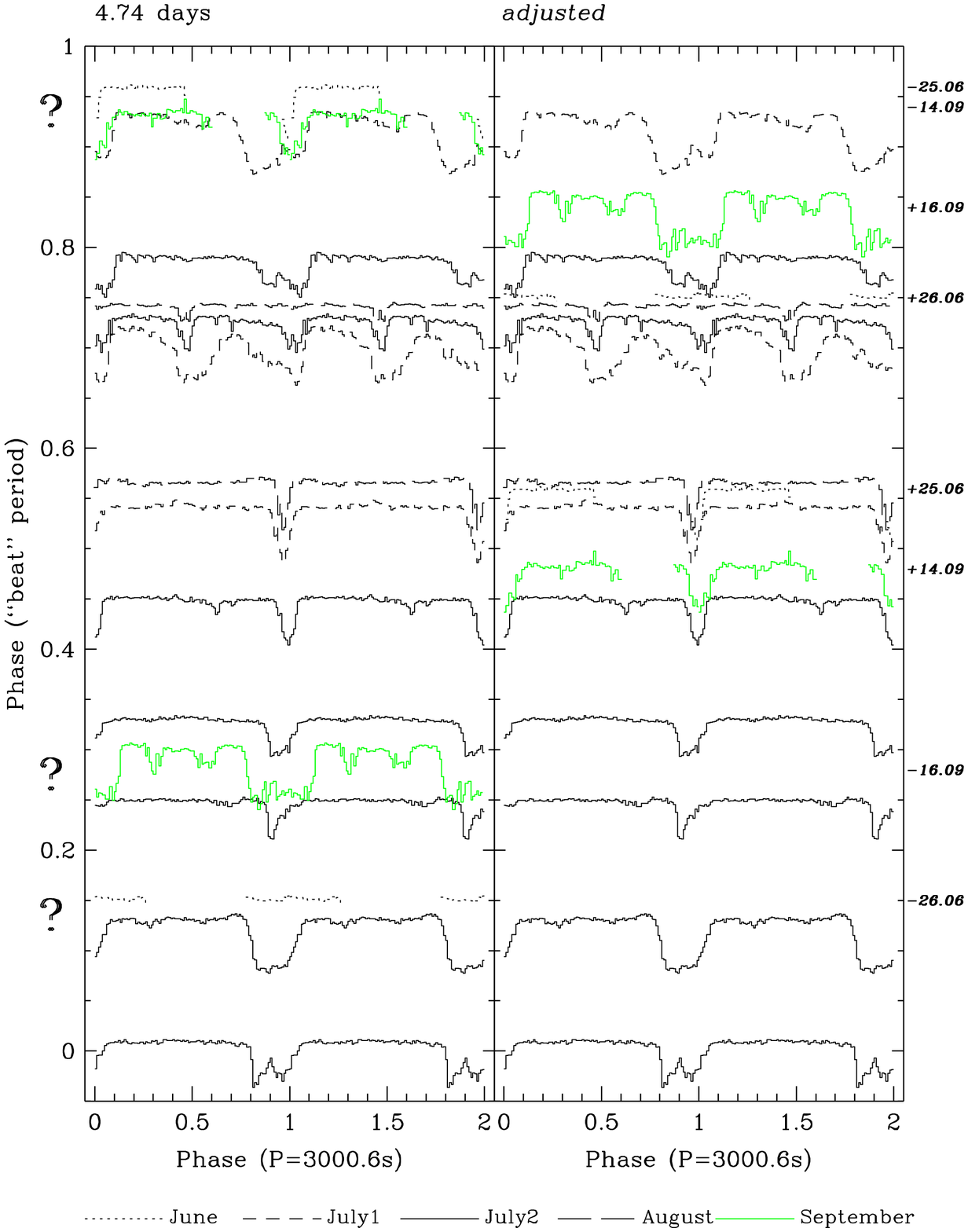}}}
      \caption{Left: each of the X-ray datasets folded on $P_X$ and normalised by dividing by the non-dip flux level, and then plotted as a
    function of phase of the 4.74\dy\
    ``beat'' cycle.  The folds corresponding to different cycles have been marked by different line styles (see key at bottom).  The `?'
    indicate cases where the phasing is inconsistent with the expected systematic morphology evolution over a ``beat'' cycle. Right: in order
    to achieve consistency, the datasets from June and September have been offset in phase by hand. The original (`--') and new (`+') phases for each
    discrepant dataset have been marked. \label{fig:trailfds1}}
    \end{figure*}

\begin{figure*}[!htb]
      \centering
\resizebox*{0.75\textwidth}{0.75\textheight}{\rotatebox{0}{\includegraphics{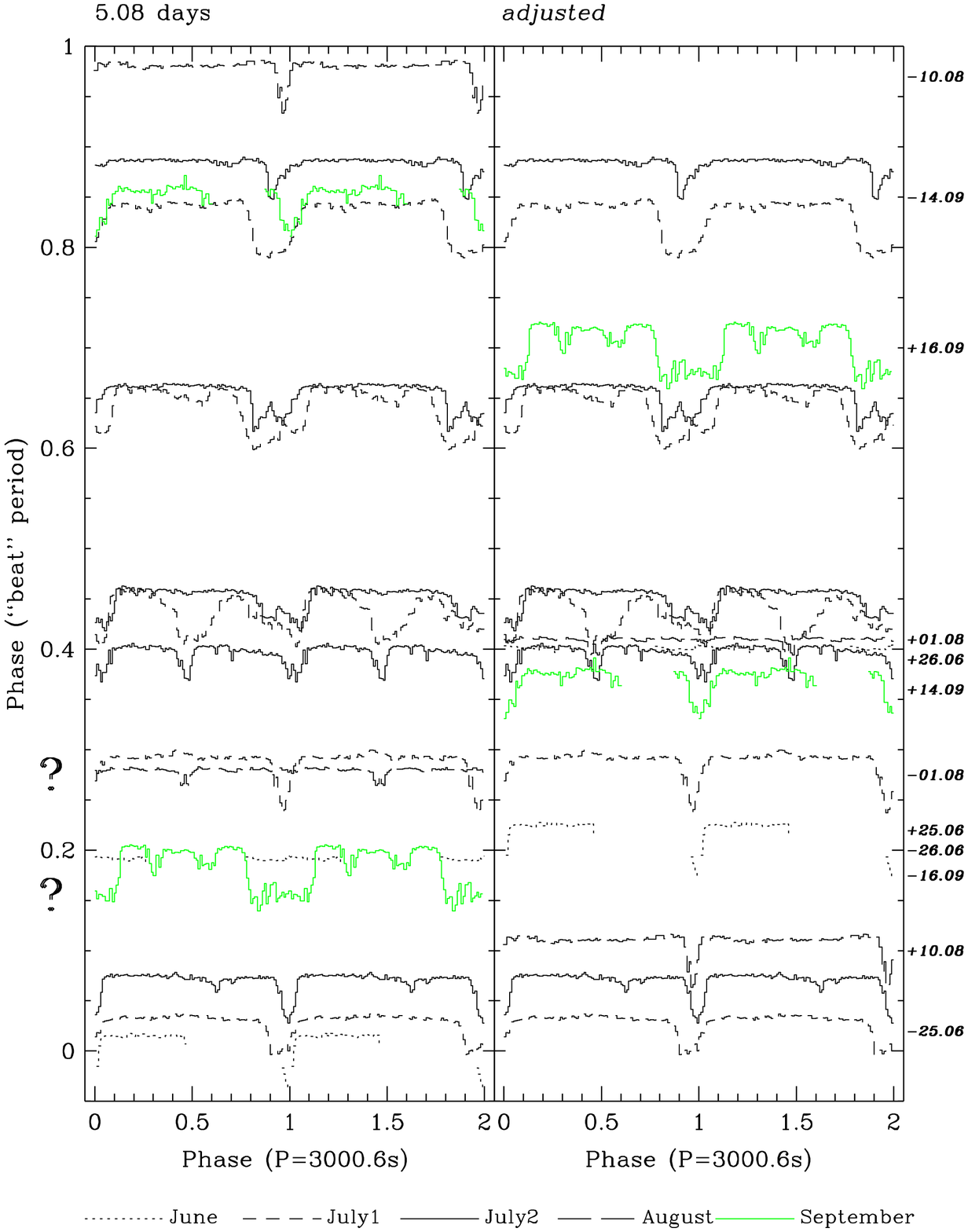}}}
      \caption{See as for figure~\ref{fig:trailfds1}, but using the 5.08\dy\ candidate ``beat'' period, and in addition the August datasets have
    also been adjusted in the right-hand panel.\label{fig:trailfds2}}
    \end{figure*}

  Turning to the ``beat'' cycle phasing
of the folded lightcurves we still find that none of the candidate periods yields a repeatable cyclic variation in
    morphology.  Only for the 5.08 and 4.74\dy\ periods, are there not inconsistencies within one ``beat'' cycle,
    or between the consecutive July cycles, as shown in the left-hand panels of figures~\ref{fig:trailfds1} and ~\ref{fig:trailfds2}.  For these two, one can then at least achieve consistency by assuming that the ``beat'' phase can change over the course of a few
    cycles, as shown in the right hand panels. In fact the assumption of such a quasi-periodic ``beat'' period would be consistent with both the clock-noise known in the
    well-studied precession periods of SS433 and Her X-1 \cite{bayk93} and the evolving superhump periods inferred in SU UMa systems during their
    super-outbursts \cite{warn95b}.  However, the shifts here are relatively large: for the 5.08 d period each month's data requires shifting relative to the others by
    between 0.1 and 0.5 in phase, whilst
    for the 4.74 d period the shifts are about 0.4 in phase, but only the extreme June and September
    datasets are discrepant with July/August.

\begin{figure*}[!htb]
      \centering
\resizebox*{.9\textwidth}{!}{\rotatebox{-90}{\includegraphics{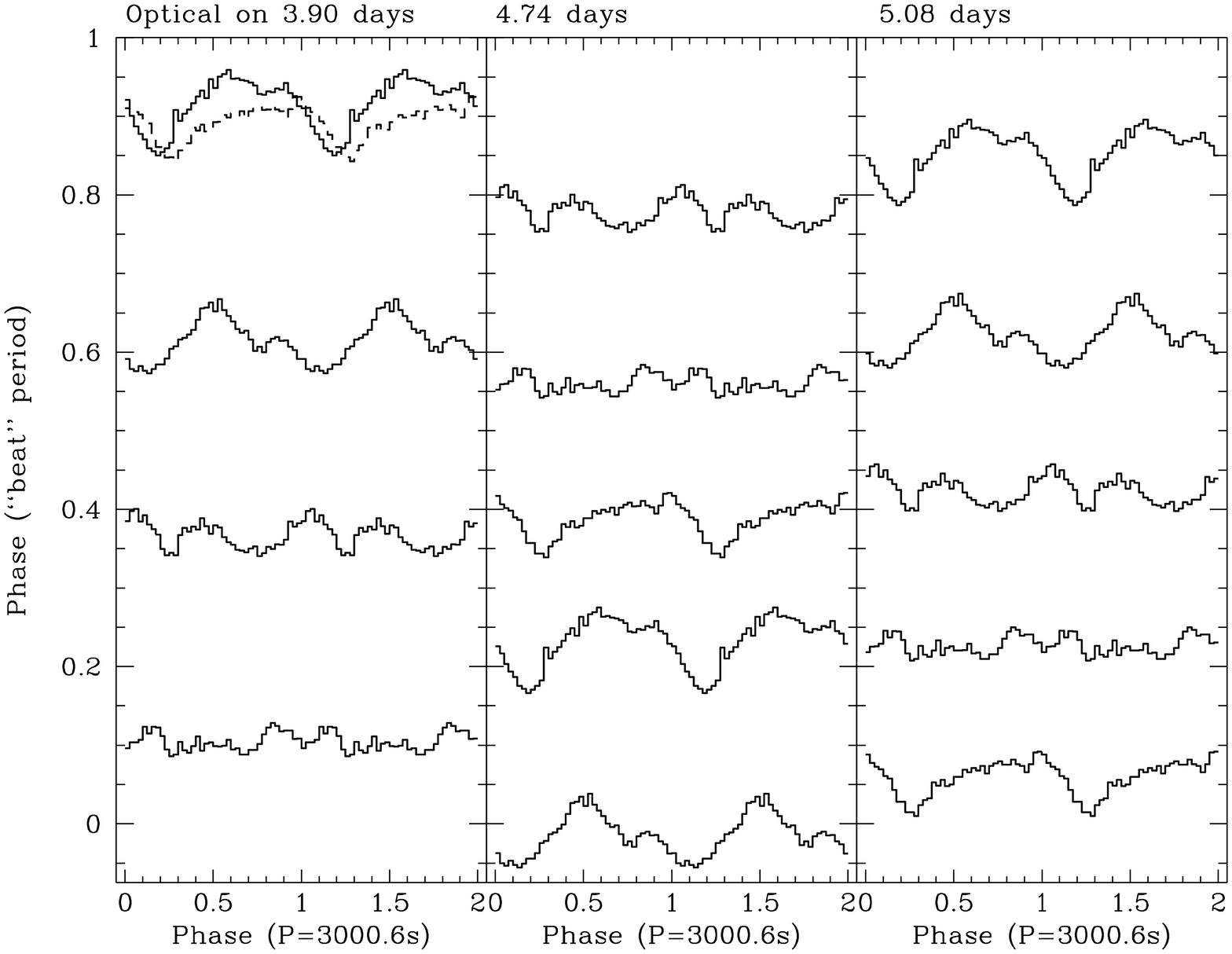}}}
      \caption{The optical data on \targ\ normalized, folded on $P_X$ and binned,  and then plotted at the
      appropriate ``beat'' cycle phasing, using the 3.90, 4.74 and 5.08\dy\ periods respectively. The ephemerides for each panel have been set so that direct
      comparison of X-ray and optical can be made.  In the latter two cases, this phasing agrees with the adjusted versions. \label{fig:trailfds_opt}}
    \end{figure*}

As a further test and to enable later comparison, we have also plotted the optical lightcurves (folded on  $P_X$) against ``beat'' cycle
phase in figure~\ref{fig:trailfds_opt}.  The phase has been set to agree with the (adjusted) phasing of the X-ray lightcurves from June.
However, we see that the optical data does not
not permit any further discrimination between candidates.

To summarise, we have found that  $P_{beat}$=3.90\dy\ is inconsistent
with a systematic, cyclic variation in morphology. We have attempted to identify other candidate periods from an analysis of the varying dip
phases, but again discrepancies are found. Only if we then relax the condition of strict periodicity
and allow phase shifts between well-separated cycles do we find that a period of $4.74\pm0.05$\dy\ can best represent the period on which the
morphology changes repeat.

\section{Longer term variations in flux levels}
\label{sect:Tanal3}
\begin{figure*}[!htb]
      \centering
      \includegraphics[scale=0.6,angle=-90]{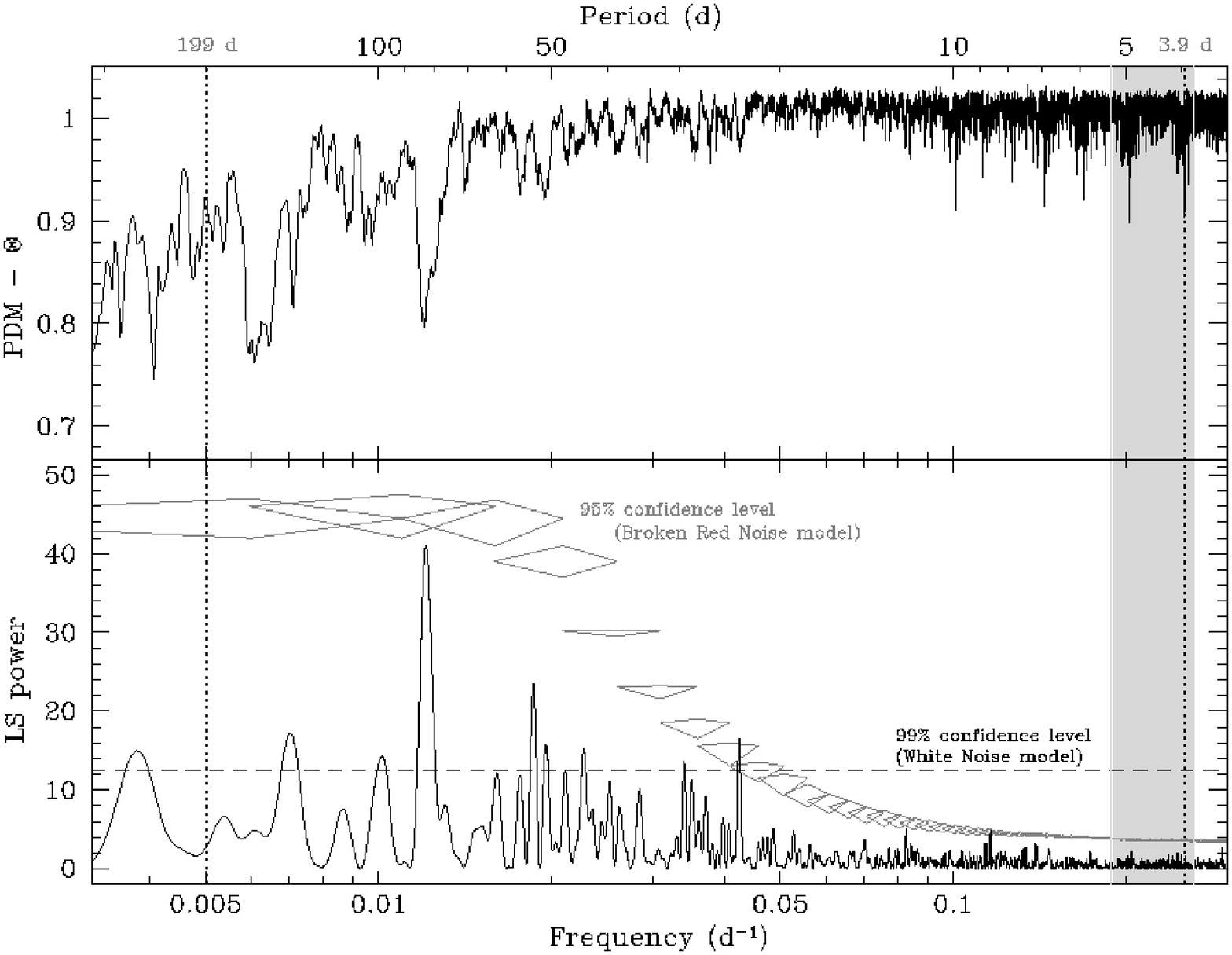}
      \caption{Phase Dispersion (upper) and Lomb-Scargle (lower) periodograms of the complete \til4 yr \sat/ASM lightcurve of \targ. A
number of peaks are present, but there are none corresponding to either the previously detected 199\dy\ modulation or the expected 3.90 ``beat'' period
(marked by dashed vertical lines) nor any even in the range of possible  ``beat'' periods (shaded region). However, those at \til0.012, \til0.043, \til0.083 and \til0.116\pd\ (\til83, \til23, \til12, \til8.6\dy\ ) are
statistically significant ($\simgt$95\% confidence) in excess of the red-noise level in their region of the power
spectrum. \label{fig:ASMper}}
    \end{figure*}

To search for any periodicities in the \sat\ /ASM data, we calculated both Lomb-Scargle (LS) and PDM periodograms from the one-day average data.
These two methods are complementary, the former modified Fourier transform being ideal for finding low amplitude, smooth/sinusoidal variations, and
the PDM, any highly non-sinusoidal periodic modulations.  The results are presented in Fig.\ref{fig:ASMper}, where the periods of interest are
marked by dotted lines/shading, namely the
previous 199\dy\ detection, the $P_{beat}'$=3.8-5.1\dy\ range and the 3.90\dy\ ``beat'' of the X-ray and optical periods. 

\begin{figure*}[!htb]
      \centering
\resizebox{.7\textwidth}{!}{\rotatebox{-90}{\includegraphics{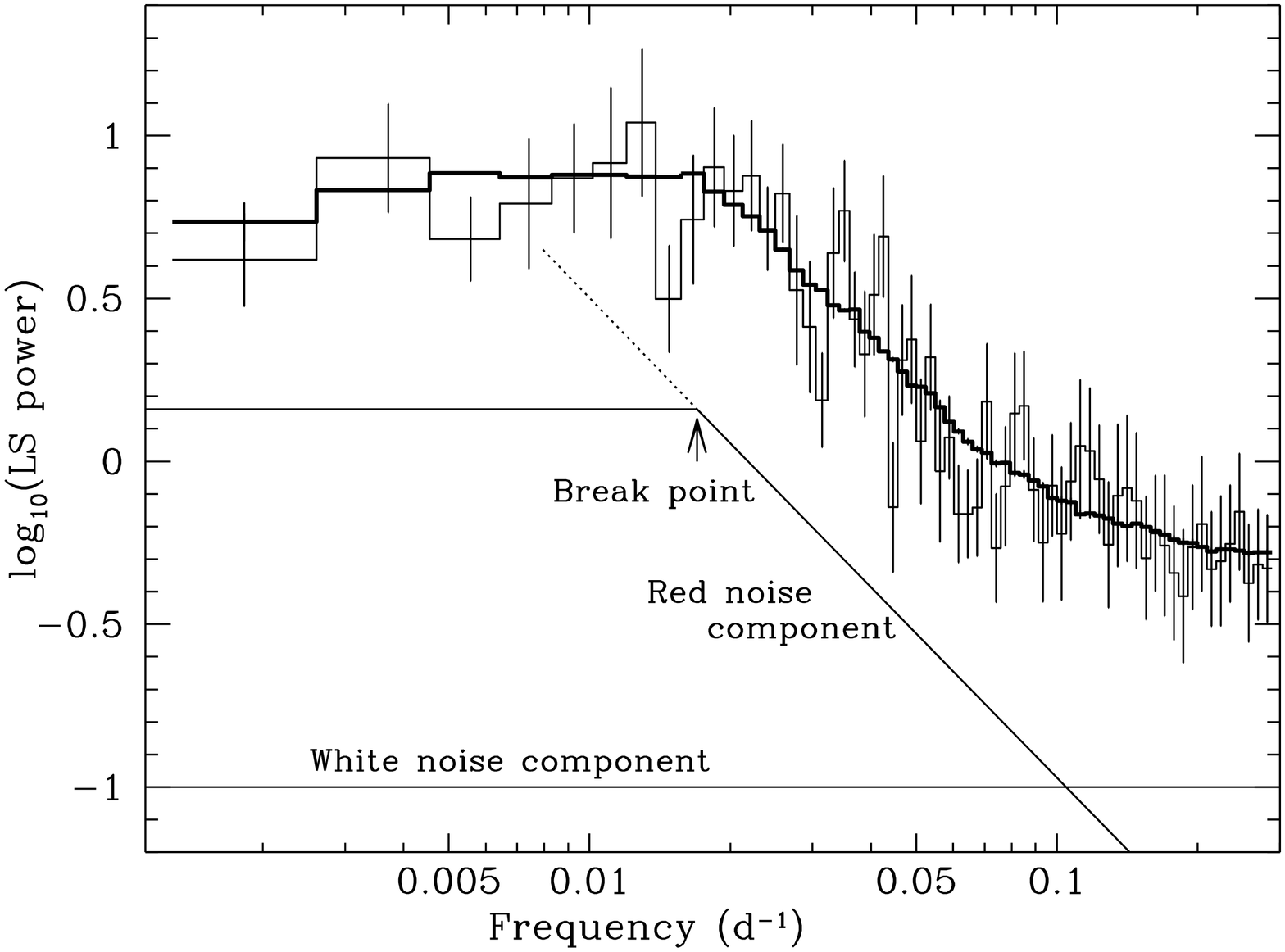}}}
      \caption{Log-Log binned LS periodogram (fine histogram, with error-bars) of complete \sat/ASM lightcurve of \targ.  Note the general form, with approximately a power
law between the break at \til0.02\pd\ and 0.1\pd\ (red noise part), which tails off to higher frequencies as the white noise contribution (flat
spectrum) becomes dominant. The components of the input model noise spectrum (white noise, plus
red noise, with a 
break at 0.02\pd\ ) are shown offset vertically and the resulting average (over 450 simulations) is over-plotted (in a thick line).\label{fig:ASMpsp}}
    \end{figure*}

However, in order to say anything meaningful regarding the detection of any variability at these or any other frequencies we must consider the
significance levels of the peaks (most reliably done for the LS periodogram).  In figure \ref{fig:ASMpsp} the log-log binned power spectrum
is shown, which clearly cannot be described by white noise, as the LS power is not approximately constant with frequency.  Rather it has the
appearance of a red noise power law,
but with a break at about 0.02\pd\ and decreasing slope at the highest
frequencies due to Poisson/instrumental white noise contributions.     

 As a consequence, the method outlined in \scite{scar82} (and implemented in
for example \pcite{home98}) for determining
significance levels in the presence of white noise cannot be applied directly in this instance.  Instead, we have developed a modified
approach, whereby an input model noise spectrum replaces the straightforward white noise.  An average over many resulting simulated noise power spectra is shown
over-plotted on the data in Fig.\ref{fig:ASMpsp}.  As the noise level clearly varies with frequency, we then split the full frequency range into
60 overlapping 0.01 \pd\ bins, for each of which we note the peak power due to noise in each simulated power spectrum.  In  this way we
can build up the cumulative distribution
function for the peak powers due to noise, and finally set a 95\% confidence level, above which any peak in the data has less than a 5\% chance
of being due to a noise component, i.e the strict periodicities which we seek.

Hence, we can see that no peaks significant above
noise are present anywhere close to the {\em a priori} periods of interest.  Setting upper limits we find that a 20\% (semi-amplitude) at 199 d, and 22\% at
\til4\dy\ signal would be detected with less than 95\% confidence.

\begin{table*}
\begin{minipage}{120mm}
\caption{Semi-amplitude of flux modulations in the \sat\ /ASM lightcurves\label{tab:Plong_amps}}
\begin{tabular}{l l c c c c } 
Energy range & Mean ASM &\multicolumn{4}{c}{Percentage peak-peak flux variation}\\
(keV)			&Count rate	(s$^{-1}$)& 83.0\dy\ & 23.4\dy\ & 12.1\dy\ & 8.6\dy\ \\
 1.3--3.0 &	0.44 &$60\pm20$ &$10\pm10$&$30\pm20$ &$50\pm20$\\
 3.0--5.0 &	0.38 &$50\pm20$ &$20\pm20$&$10\pm10$&$10\pm10$\\
 5.0--12.1 & 0.54 &$50\pm20$&$50\pm20$&$20\pm20$&$20\pm10$\\
1.3--12.1& 1.16 &$60\pm10$&$40\pm10$&$20\pm10$&$20\pm10$ \\
\end{tabular}

\end{minipage}
\end{table*}  

However, from the LS periodogram moderately significant ($\simgt95\%$ confidence) periodicities are easily identifiable at about 0.012 \pd\, corresponding to  $83.0\pm0.4$\dy\, and at
\til0.043\pd\ (23.4\dy\ ), and with smaller amplitudes at $\sim$0.083 and $\sim$0.116\pd\ ($\sim$12, $\sim$8.6\dy\ ), although only the first of
these is evident in the PDM.  Indeed, the 83.0\dy\ period can be clearly seen in the binned lightcurve itself, as indicated in
figure~\ref{fig:obslog}. Consideration of the times of maximum flux also imply that it is not strictly periodic.  

\begin{figure*}[!htb]
      \centering
\resizebox*{.8\textwidth}{0.4\textheight}{\rotatebox{-90}{\includegraphics{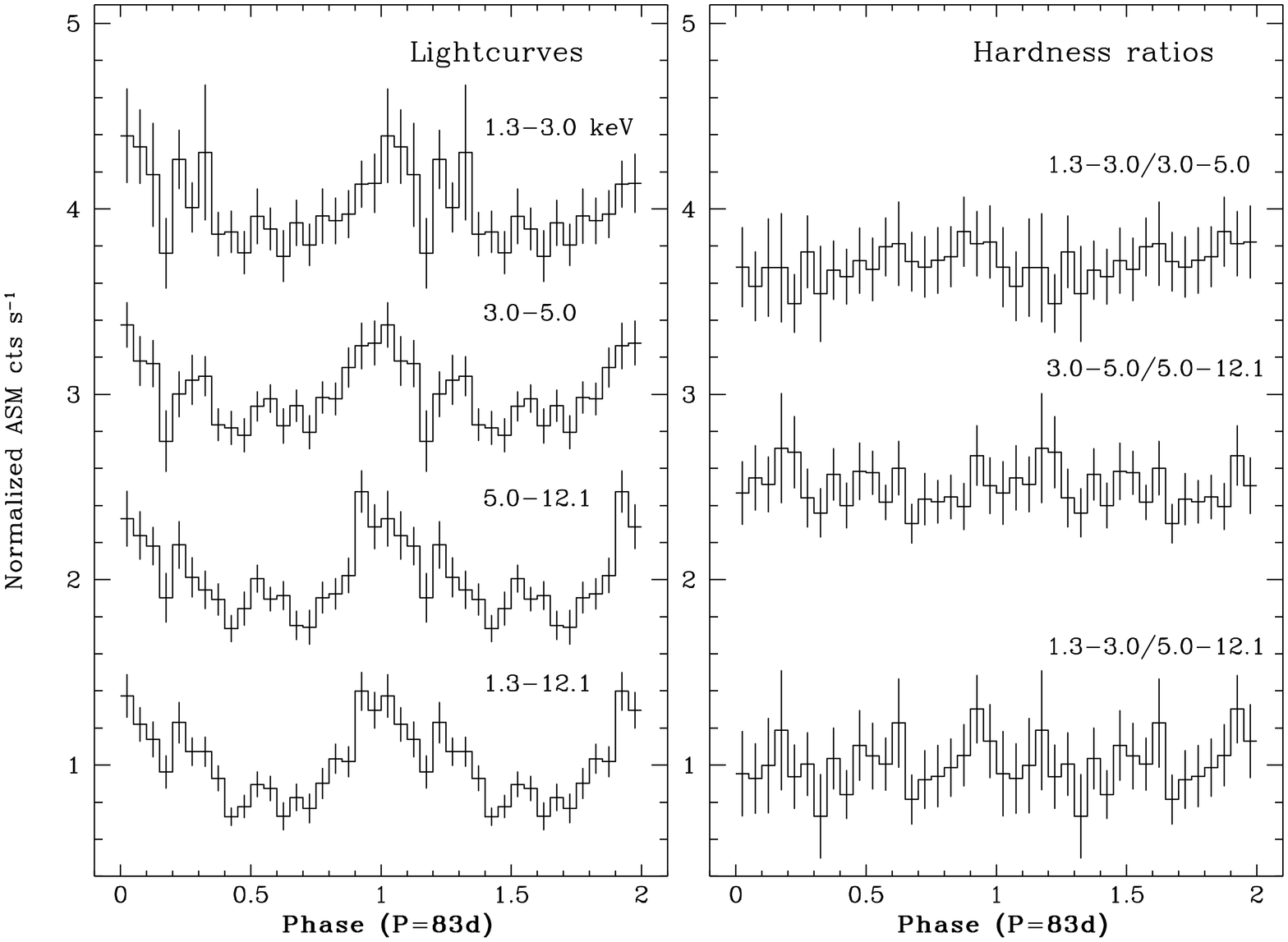}}}
      \caption{Left: One-day averaged \sat/ASM data on \targ\ in three energy bands and complete folded and phase binned on an 83\dy\
    period.  The lightcurves have been normalised and then offset vertically (by 1 cts s$^{-1}$) for clarity.
    Right: hardness ratios between the different bands clearly demonstrate the approximate energy independence of the modulation amplitude at this
    period (right).\label{fig:ASMflds}}
    \end{figure*}

To shed further light on all the detected periodicities, we also considered the X-ray data from the complete band and in three energy bands folded and phase binned on each.  The estimated peak-to-peak
amplitudes for all modulations are listed in table~\ref{tab:Plong_amps}.  For the three shorter periods, the amplitudes vary in a
non-systematic manner with energy, hence casting doubt upon their reality.  The signal-to-noise is simply too poor for such low amplitude
effects. Only in the case of the 83\dy\ signal do we see a significant modulation in each of the bands, although we note that its amplitude is much
less than that of the supposed 199\dy\ signal found in {\it Vela 5B} data and hence it would not then have been detectable.   Notably, the
amplitude is approximately the same in each band, indicating energy independence (as confirmed by the hardness ratios, see Fig.\ref{fig:ASMflds}).  Both the
smooth morphology and lack of significant energy dependence, argue against an obscuration effect such as that seen due to the precessing disc
in Her X-1 \cite{gere76,cros80}. It seems more likely that \targ\ is similar to X1820-303, in which the X-ray flux shows a 176\dy\ periodicity, also
correlated with bursting activity, and hence an inherent physical property of the system, most likely due to a variation in the accretion rate
(see also \pcite{blos00}).

\section{Optical periodicities} 
\label{sect:Tanal4}
The optical data exhibits significant (0.5 mag) variations from night to night over the five consecutive runs (see Fig.~\ref{fig:obslog} lower panel).  Although larger in amplitude,
they are consistent with
the changes in the X-ray flux, as expected.  In any case, for consideration of the \til3000~s
timescales, of which there are at least 2.5 cycles per night, we can safely detrend each night's data with a linear polynomial fit, removing
any longer term effects.
   \begin{figure*}[!htb]
      \centering
      \includegraphics[scale=0.6,angle=-90]{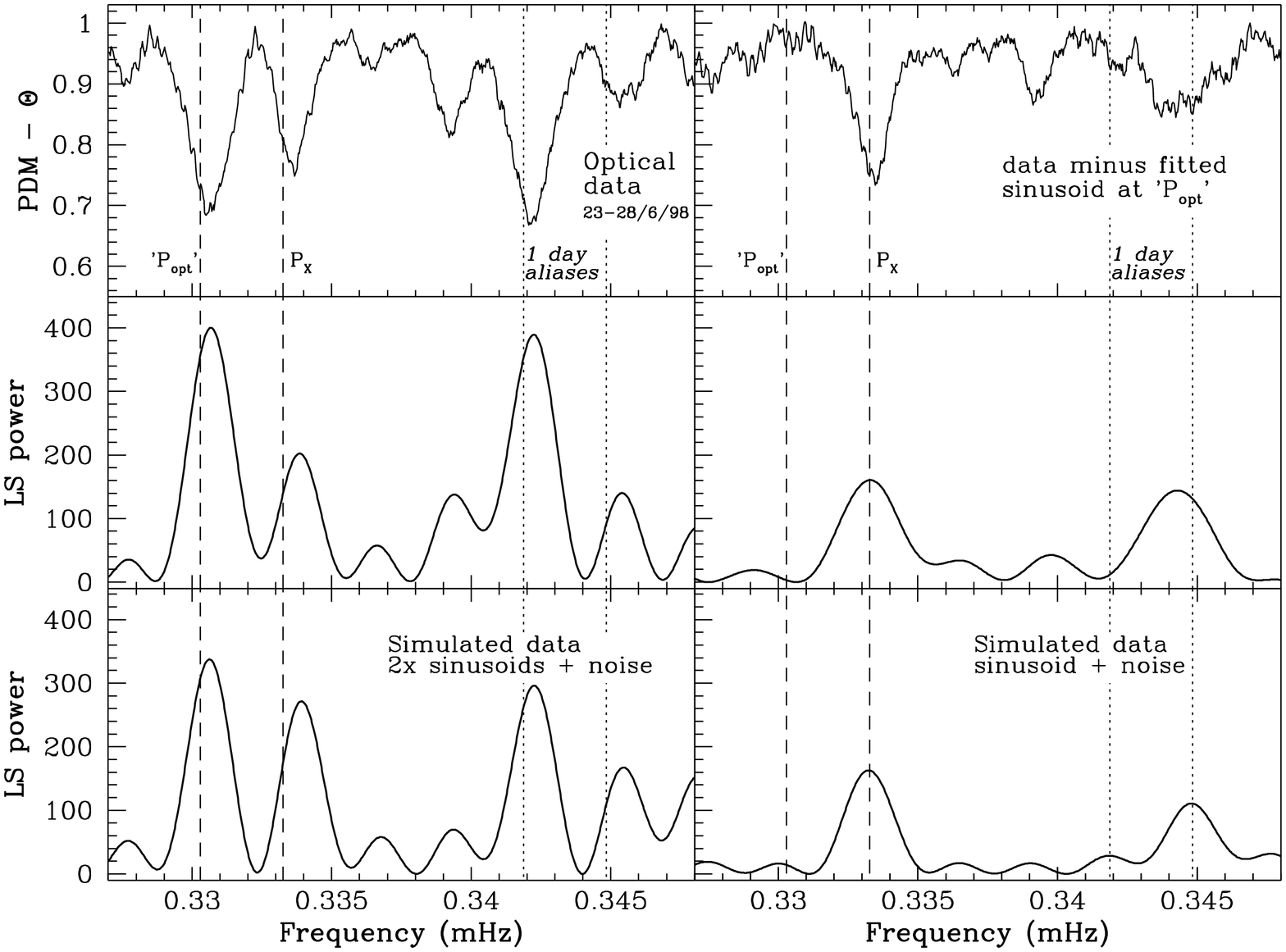}
      \caption{Phase Dispersion (upper) and Lomb-Scargle (middle) periodograms of the complete 5 night optical lightcurve of \targ\ (left) and
the same data after subtraction of the measured sinusoidal ``orbital'' modulation at 3024.2 s (right).  The 1 day aliasing due to the
sampling complicates the periodograms, but as shown with the simulated data (bottom) the structure is reproducible simply by supposing there
are two signals at close to $P_X$ and $P_{opt}$.  Note: (left) the
measured peaks are shifted relative to the known
$P_X=3000.6$\sec\ and $P_{opt}=3027.6$\sec\ as indicated by the vertical lines (dashed and dotted), whilst after the removal of the latter
(right) the
measured $P_X$ agrees with our X-ray determination. \label{fig:NOTpgm}}
    \end{figure*}

Once again, both an LS and PDM periodogram were calculated from this detrended lightcurve (see Fig~\ref{fig:NOTpgm}).  There are a number of
significant peaks apparent in both periodograms.  Of course the one-day aliasing due to the sampling can cause confusion, but with {\it a
priori} knowledge of which frequencies to consider, we may identify the principal periodicities quite easily.  The largest peak
($\nu=28.570$\pd\, indicated along with its one day aliases) corresponds to
$P=3024.2\pm0.4$~s, close, {\it but not consistent} with the 3027.6~s optical periodicity of Callanan, Grindlay \& Cool (1995, henceforth $P_{opt}$).  The apparent discrepancy
of 8$\sigma$ is of note; it may simply indicate the underestimation of the true uncertainty in the peak frequency, due perhaps to
the effects of the larger amplitude (and varying) modulation present on the nearby X-ray period, or indeed may be evidence for a period change.
We will consider the stability of the optical period further in the next section.  Excluding these,
the next most significant peak is indeed at $2995.8\pm0.6$\sec\ (28.84~\pd\ ), which is close to the X-ray period.  Subtraction of the
longer period modulation reveals the signal on $P_X$ even more clearly.  Notably, this has a revised period of  $3000.2\pm0.6$\sec\ consistent with our X-ray determination.  Moreover, this signal appears relatively more distinct
in the PDMs, implying that it is non-sinusoidal (as one would expect).  Such an occurrence of both periodicities in the optical was also seen in data
from 1990 September by \scite{call95}, when they comment that \targ\ was in an anomalously low optical state, and by \scite{grin89}.

As we did for the X-ray data, we then proceeded to examine the folded and binned lightcurves for each day and the complete dataset, using both
the period and ephemeris from \scite{call95} and our own X-ray period.  The overall  lightcurve folded on $P_{opt}$ shows an almost sinusoidal
profile, but with a much smaller amplitude (\til10\%), than the nightly folds (up to \til25\%).  \begin{figure*}[!htb]
      \centering
      \resizebox*{1.0\textwidth}{0.4\textheight}{\rotatebox{0}{\includegraphics{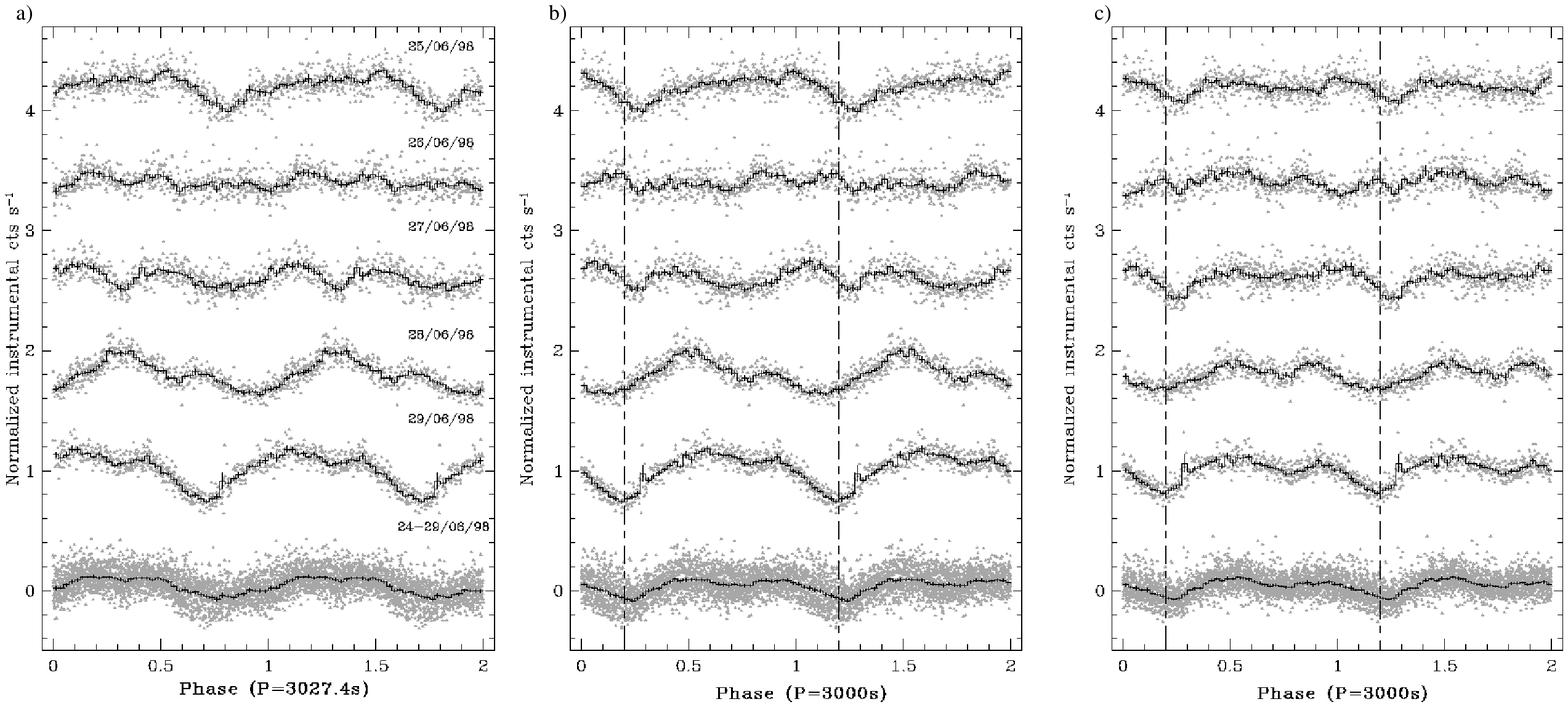}}}
      \caption{a) The optical lightcurves from each night and overall folded on $P_{opt}$ (grey points) and phase binned (over-plotted in
black).  The data was firstly 
normalised by dividing by the mean instrumental flux value, and they have been plotted vertically offset for clarity. Note that the nightly
modulations are varied in morphology and share no phase relation to each other or to the overall fold, which is also much more smoothly sinusoidal
in appearance. b) Same as (a) except that the
3000.6s X-ray period has been used for folding.  Now the clearest features of the modulation appear at the same phase, indicating that this shorter period is
the dominant, over the course of a few cycles at least  (c) Before folding on
$P_X$, the sinusoidal fit at $P_{opt}$ (good approximation) to the complete optical dataset has been subtracted.  The features which must be
related to the X-ray dips are still clearly present.
\label{fig:optfolds} }
    \end{figure*}
In common with the X-ray folds,
there is clear evolution of the morphology from day-to-day.  In particular the phase of the principal minimum appears to shift systematically
relative to the plotted sinusoid.  If, indeed, the 3027s period represents that of the binary orbit, so long as the reprocessing geometry is in
some way asymmetric, the change in its aspect will lead to the smooth modulation we see.  Hence, we
fitted and subtracted a sinusoid fit to the overall fold (as a good approximation to its shape), to leave the effects presumably due to obscuring disc
structure alone.  The folds in Fig.~\ref{fig:optfolds} shed further light.  It is now clear that
the {\it remaining} structure in the optical lightcurve varies on $P_X$ not $P_{opt}$  This strongly infers that the
same effect causing the X-ray dips, creates corresponding optical `dips'.  Comparing X-ray and optical folds (on $P_X$) from the same phase of
the ``beat'' cycle (Figs.~\ref{fig:trailfds3.9}--\ref{fig:trailfds_opt}), we see that there is surprisingly little evidence of a clear relationship between the two.  For
instance, in the simultaneous data of June 25 the X-ray exhibits a distinct narrow dip, whilst there is very little optical modulation at all.  Nevertheless, we
may determine the phase relationship of the dips in each band in the other cases, finding that the optical dip lags the X-ray by
\til0.2-0.3 in phase. 
\begin{figure*}[!htb]
      \centering
      \includegraphics[scale=0.61,angle=-90]{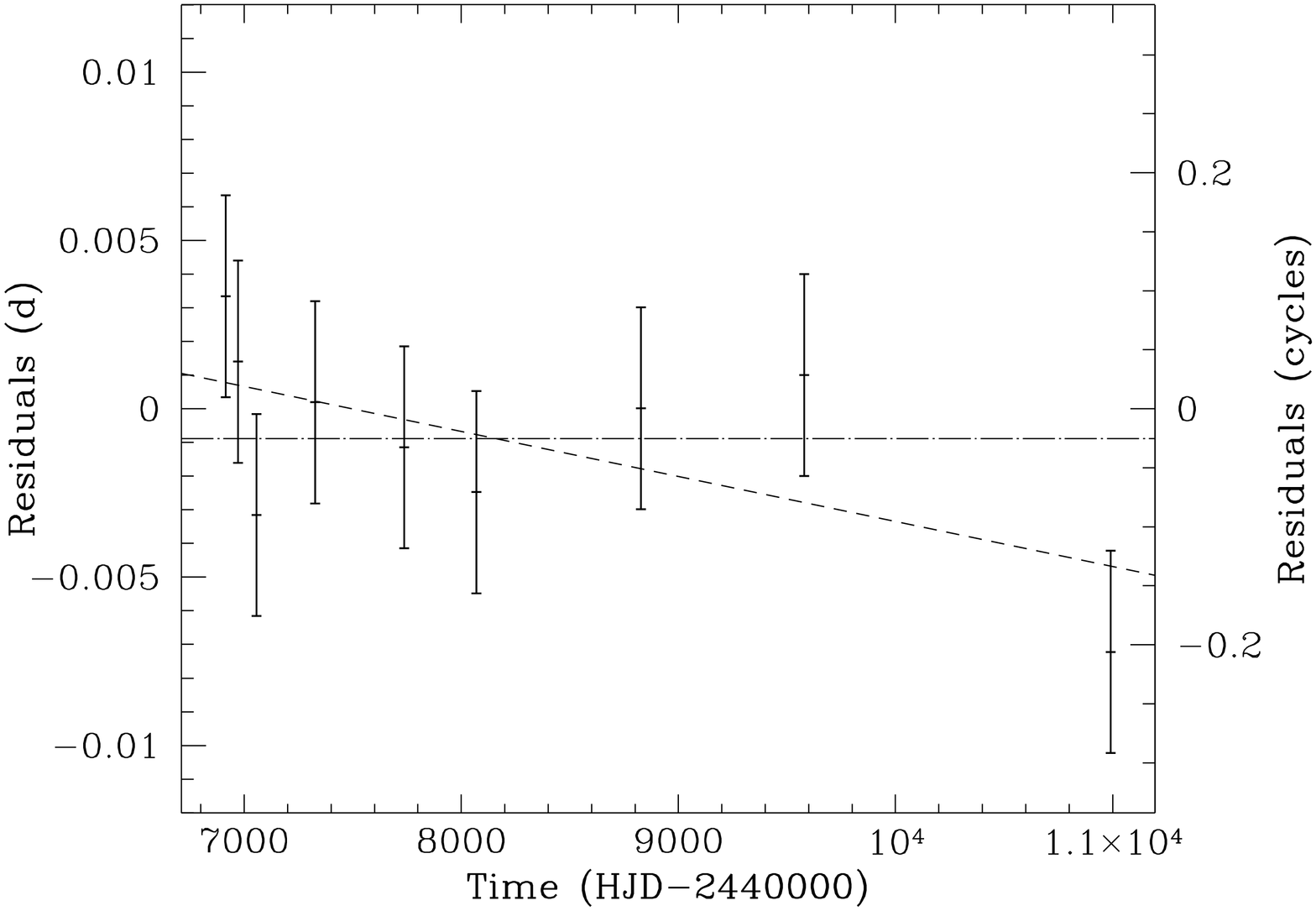}
      \caption{Using the Callanan, Grindlay \& Cool (1995) optical period the residuals between the predicted time of optical minimum and that measured is plotted
for all the observations 1987 through 1998, together with a constant fit, as appropriate for their ephemeris.  Clearly, our 1998 point has quite
a large residual implying the possible need for a longer period
(as indicated by the linear fit).  Nonetheless, note how small the scatter appears over the entire 11 year baseline, inferring its high stability. \label{fig:optephem}}
    \end{figure*}
 
\subsection{Revised optical ephemeris}
\label{sect:optephem}
We may also reconsider the optical ephemeris derived by \scite{call95}, by using the complete optical lightcurve folded on their ephemeris (see Fig.~\ref{fig:optfolds}a
lowest folded lightcurve) to extend the baseline by a
further 4 years.  Clearly, the minimum is shifted relative to phase zero.  Applying their method we fitted a Gaussian to the principal
minimum giving $\phi_{min}=-0.20\pm0.09$.  The {\it a priori} probability of obtaining a minimum this close to the published ephemeris
(equivalent to \til1$\sigma$) is still \til40\%.  Hence, the period could have changed significantly since the earlier observations, as
suggested by 
the periodogram result, and we simply have a chance approximate alignment.  On the other hand, if we assume that the period is stable
then the shift implies that the 1995 ephemeris is slightly incorrect.
  So to update it, we also digitised the residual values as
presented in \scite{call95} Fig. 6 (but with the corresponding HJD values calculated from the quoted observation time/dates).  In
figure~\ref{fig:optephem} we present all the residuals to $P_{opt}=3027.551$~s from 1987 through 1998.  If this period is correct then a
constant fit will be acceptable, whereas a slightly different period will be represented by a linear term.   For the constant fit
we find an acceptable $\chi^2_\nu=1.06$, but with a revised $T_0({\tiny \rm HJD})=244,6900.0019\pm0.001$, whilst for a new period and
ephemeris of $T_0({\tiny \rm HJD})=2446900.0128\pm0.0055+n*(0.03504115\pm0.00000002)$  ($P_{opt}'=3027.555$~s) we have a smaller $\chi^2_\nu=0.77$.
Applying a one-sided F-test \cite{bevi92} confirms that the extra linear term does provide a marginally better fit than the constant with 91\%
confidence.
However, the revised period is only 1$\sigma$ larger than the \scite{call95} result, hence not too much weight should be attached to this minor
revision.  The most important aspect of this
new ephemeris is that our data, on face value,  is consistent with all previous observations, that is with the stability of the optical modulation over more than a
decade. If confirmed this would provide yet
stronger support for its orbital origin, but still further observations are needed to do so, preferably spaced by a few months to maximise the accuracy
of the cycle count.


\subsection{Further comparison of optical and X-ray modulations}
\begin{figure*}[!htb]
      \centering
      \resizebox*{1.0\textwidth}{0.5\textheight}{\rotatebox{-90}{\includegraphics{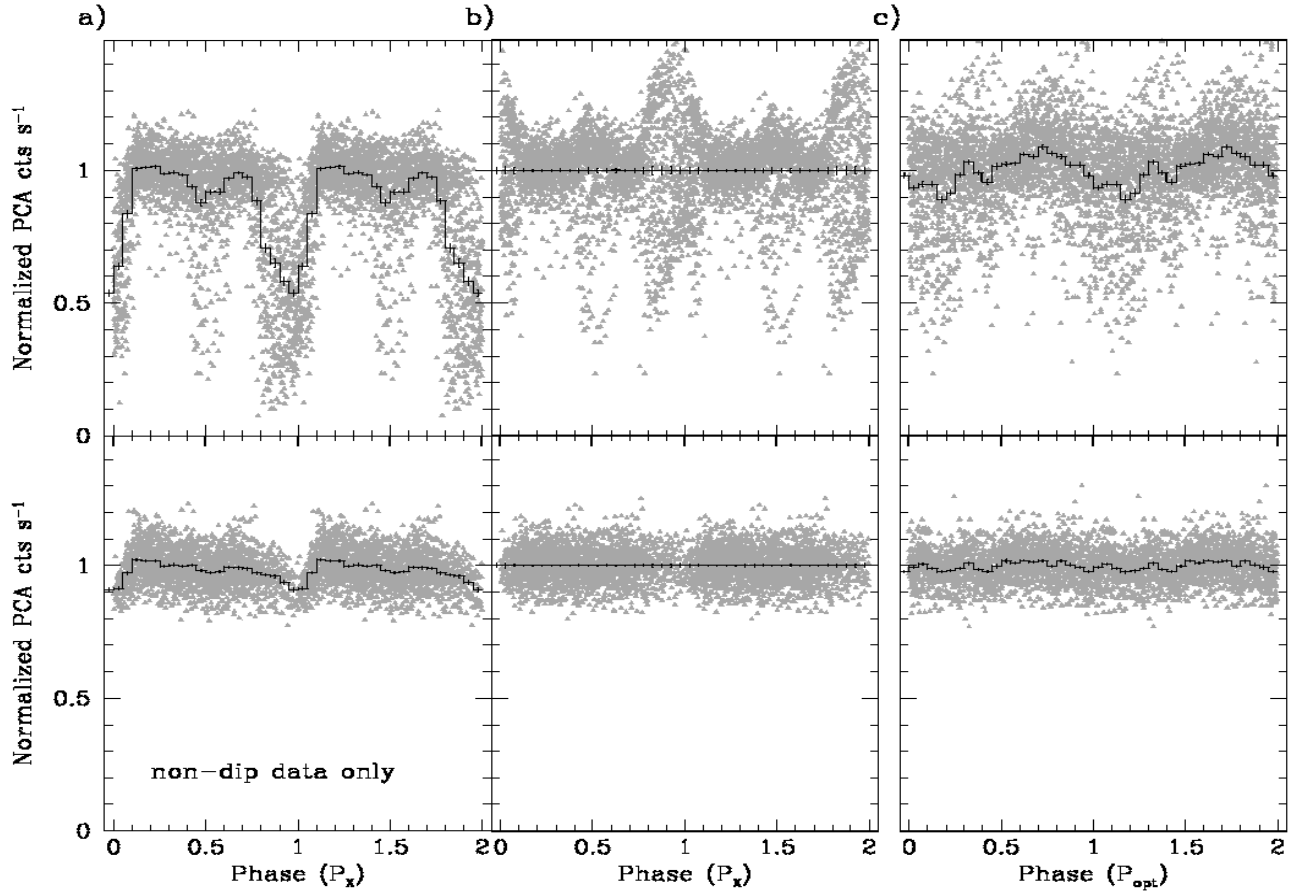}}}
      \caption{Upper panel:  Complete June-September \sat/PCA lightcurve on \targ . Prior to folding, each of the daily datasets was normalised
      by dividing by the non-dip flux level.  a) Lightcurve folded and phase binned on the X-ray period of 3000\sec . b) After
      subtraction of this average (binned) modulation from the lightcurve, folded once more on $P_X$ (offset by 1 for clarity). c) Same
      subtracted data as in (b), but now
      folded and phase binned on the {\em optical} period of 3027\sec\ (offset by 1 for clarity).   An apparent modulation on $P_{opt}$ remains even after subtraction of
      the mean X-ray modulation shown in (a).  Lower panel: Only the non-dip data (i.e. applying a 75\% flux
      cut to remove the dip data points).  Without the dip data there is no longer any modulation on $P_{opt}$, showing that this only arises
      from the uneven sampling of the dips. \label{fig:xfolds2}}
    \end{figure*}

Having considered the optical lightcurve folded on the X-ray period, we will now look at the converse.  In figure~\ref{fig:xfolds2} we present
the complete X-ray lightcurves folded (and phase binned) on both $P_X$ and $P_{opt}$.  The presence of both primary and secondary dips becomes
clear in the $P_X$ fold, showing how the latter can be as deep, but only occur a fraction of the time.  Before folding on $P_{opt}$ we subtracted
the mean modulation on $P_X$ (i.e. the binned fold) in the same way we treated the optical data. When folding on this slightly longer period,
the primary dips are spread over a much larger phase region, and are only limited due to the effects of sampling; leading to the apparent
modulation in the phase binned plot.  Lastly, considering only the non-dip data we see that there is in fact no evidence of any overall
modulation unrelated to the dip behaviour.

\begin{sidewaystable}
\caption{Summary of our results on the temporal phenomena of \targ\ \label{tab:summary}}
\begin{tabular}{l l c l } 
Temporal phenomena & Analysis undertaken & Result & Notes\\ 
Recurring X-ray dips &PDM of entire X-ray (PCA) dataset & $P_X=3001.0\pm4.0\sec$ & Drops in flux from non-dip level of varying \\
								&	Cycle counting between 6 narrow & & duration, but all within narrow phase range.\\
								&	 dips (over 3 month span) &$P_X=3000.6\pm0.2\sec$  &\\
Smooth modulation & Cycle counting between minima& 	$P_{opt}=3027.555\pm0.002\sec$&  Smooth, almost sinusoidal variation\\
of optical flux						& over 1987--1998&&but on short timescales\\
&&&dominant features not phase stable\\
&&& (follow X-ray period  instead)\\
``Beat'' period & $P_{beat}=(P_X^{-1}-P_{opt}^{-1})^{-1}$  & $P_{beat}=3.90\pm0.03\dy$ & - \\
Dip morphology cycle & Sinusoid fitting to dip phase variation + &  $P_{beat}'=4.74\pm0.04\dy$& Dip phase variations approximated by sinusoid,\\
										&	consistency with systematic/cyclic evolution&&but evidence for quasi-periodic nature\\
Longer-term flux& LSP/PDM of 1 day average &	 $P_{long}=83.0\pm0.4\dy$ & Smooth variation (though non-sinusoidal)\\
 modulation 												&	X-ray (ASM) flux levels	&& Energy independent amplitude\\	
	
\end{tabular}

\end{sidewaystable}  
\newpage
\section{Discussion}
\label{sect:disc}
Taking a  qualitative approach we will consider the implications of our results.  A range of models have previously been put forward
to explain the observational properties of \targ\, ranging from a hierarchical triple system, to something
akin to the Cataclysmic Variable (CV) superhumping systems with their precessing discs.  We also outline a modification to the latter, related to the more recent
discovery of negative superhumps. Dealing with each model in turn, we will confront its predictions with our detailed observational results.

\subsection{Hierarchical triple}
\scite{grin88b} and \scite{grin89} proposed that \targ\ is part of a hierarchical
triple system in order to provide an explanation for the two periodicities close to 3000\sec\ and the then reported longer-term 199\dy\ flux variation.  In their model, the
ultra-compact 3000\sec\ binary, whose orbital period is that measured in the optical, is orbited by a third star on a somewhat
longer, retrograde orbit ($P_{orb}\simeq4$\dy).  The gravitational influence of this
third star leads to perturbations of the mass transfer rate on the X-ray period.  Unlike other dippers, they require
that the dips arise when the mass transfer rate is periodically enhanced, thereby forming extended disc-edge structure which then circularises (around
the disc) within one binary orbit. In this way, the problem of the dip-causing structure being fixed in binary phase is solved.  Moreover, the
long-term flux modulation is also explained, as arising from the slow changes in binary eccentricity at the period $P_{long}\sim
P^2_{outer}/P_{inner}$ \cite{maze79}, where $P_{inner}$ refers to the inner binary orbit and $P_{outer}$ to the outer retrograde orbit of the third star. 

However, one must bear in mind, before even considering the observational constraints, that the formation of such a hierarchical
triple (with a retrograde third star) is only probable within the crowded environment of a globular cluster core.  Indeed \scite{grin89} suggests that this may have been the
case for \targ.

\subsubsection{Comparison with observations}
{\it In favour:}\begin{enumerate}
\item $P_{opt}=P_{orb}$ in agreement with the apparent high stability of the overall optical modulation.
\end{enumerate}
{\it Against:}\begin{enumerate}
\item	There is no reason to suppose \targ\  has a different dip formation mechanism to the other dippers and hence we might expect some evidence
for X-ray dips at $P_{orb}$, in addition to the dips occurring at times when the mass transfer is enhanced (unless an extremely specific
orbital inclination is chosen so that orbital effects are never visible). 
\item There is no confirmation of the old 199\dy\ period, and the new 83\dy\ value is possibly too short to fit in with changes in eccentricity
driven by the third star. Following \scite{grin88} we have $P_{long}=K\times P_{outer}^2/P_{inner}$, where now $P_{long}=83\dy$, $P_{inner}=0.035\dy$ and $P_{outer}\sim\frac{2}{3}\times P_{beat}\sim\frac{2}{3}\times3.90\sim2.6\dy$,
yielding $K=0.4$ compared to their 0.7 for a constant which is expected to be of order unity.
\end{enumerate}

\subsection{Permanent superhumper}
The presence of doubly-periodic optical modulations is a well-known phenomenon within the field of CVs.  The effect was
first identified during the super-outburst events of the SU UMa sub-class of CVs, and hence the non-orbital modulations were referred to as
superhumps \cite{vogt84,warn75}.  Further study of CV lightcurves led to the discovery of permanent superhumping systems, amongst the nova-like
and nova remnant classes
\cite{warn95c}.  In all cases, the systems have extreme mass ratios, which in turn leads to the tidal distortion of the disc by the donor \cite{whit88,whit91}.  The
elliptical disc then precesses in a prograde manner  (in the inertial frame) on a timescale of days.  Both \scite{whit89} and \scite{smal92b}
have proposed that \targ\ is an
example of an analogue LMXB system, although the authors present differing ideas as to the origin of the optical modulation.  \scite{whit89} invoked a purely geometrical effect, whereby the projected area of the irradiated disc (the principal optical emitter in
the system) varies as the disc precesses.  However, the successful \scite{whit88} model for superhumps shows that such a variation would be on
the precession period, not on its beat with the orbital.  \scite{smal92b} presented a modified approach, in which the vertical extent of the
pressure-supported disc would vary on $P_{superhump}$, in response to the varying vertical component of gravity due to the secondary.  Hence,
in turn the area available to reprocess X-ray emission into optical would also vary, leading to the optical modulation we observe (also see \pcite{odon96}).
Indeed, it has been proposed that this is the analogous mechanism that works for CVs, except that it is the increased reprocessing of
ultraviolet emission from the innermost disc which provides the superhump light source (\pcite{harl92}, see also \pcite{bill96}).  Previously, the direct enhancement of viscous
energy release in the outer disc had been assumed to provide the light source \cite{whit88}.  In any
case, such superhump periods will always be longer than the binary period (by a few
percent).  Hence, in contrast to the triple model it requires the X-ray period to be that of the binary  and the optical due to the
superhump effect.\\ 

\subsubsection{Comparison with observations}
{\it In favour:}\begin{enumerate}
\item With $P_{X}=P_{orb}$ the X-ray obscuring material is straightforwardly fixed within the binary frame, and the same mechanism can work as for other X-ray dippers.  The dominant optical modulation on this period is also readily explained.  Both the vertical structure of the
impact region and the remainder of the disc will be emitting optically due to reprocessing of intercepted X-rays, and hence obscuration effects and
changes in the aspect of the disc structure can account for that modulation.
\item Presuming this material to be related to the stream impact region at the disc edge, the presence of an elliptical disc provides a ready explanation
for the smooth variation in primary dip phase over the precession cycle.
\item From previous observations, the longer period optical modulation dominates over the variations on $P_X$ when the system is brighter, and
indeed this is exactly what we would expect if the
superhump light source is due to irradiation (in contrast to the comment of \pcite{call95})
\end{enumerate}
{\it Against:}\vspace{-.29cm}\begin{enumerate}
\item The very distinct dip morphology changes, and especially the disappearance of primary dips altogether is difficult to explain simply with
this model, but this is probably not a major drawback as more detailed modelling is required.
\item Ongoing work on the permanent CV superhumpers does imply that even in these system with apparently stable mass transfer and hence disc
size (c.f. the outbursting SU UMas) the superhump periods are almost stable over the course of weeks, but not years.  This poses a problem with
interpreting the very stable optical modulation as some sort of superhump, i.e. with putting $P_{opt}=P_{superhump}$, given the very long
timescale of apparent stability of \targ. 
\end{enumerate}

\subsection{New model: permanent {\it negative} superhumper} 
In addition to the CV superhumps with periods longer than orbital, (positive), there is now evidence for such modulations with periods a few
 percent {\it shorter}, known as negative superhumps \cite{patt93}.  These often appear together with the positive kind, and in fact the proposed
 explanation is essentially the same, except that one now requires retrograde precession of some feature associated with the disc (e.g. the line of nodes).  However, this will only in general arise if the disc can be made to lie
 outside of the orbital plane, i.e. it is tilted and/or warped, which is the main difficulty.  In the case of LMXBs, we do at least have some idea of
 how to form such stable warped discs.  Irradiation-driven coronal winds have been put forward by \scite{scha94} in the case of Her X-1, where
 as soon as the disc is perturbed out of the plane, the torque on material leaving the disc can act to enhance and then preserve the warp.  We note
 that \targ\ does indeed possess an accretion disc corona, supporting the presence of such a mechanism.  Another possibility could be magnetic effects, as posed by \scite{lai99}, although
 this is only likely to be of importance in more highly magnetic systems and in any case close to the magnetosphere, i.e. the innermost disc
 regions.  Unfortunately, further work by \scite{ogil00} on the application of irradiation-driven warping to X-ray binaries has convincingly shown
 that this will not work for most LMXBs and certainly not for such an ultra-compact system as \targ.  Nevertheless, as in the case of CVs let
 us postulate that \targ\ possesses a warped, retrograde, precessing disc, and see what this model will do
for us.  

Considering the optical emission, the underlying modulation (on $P_{opt}$) will arise from some asymmetry in the reprocessing geometry, which
then leads to a contribution which varies with binary phase as the aspect changes, for instance the heated face of the secondary.  However,
\targ\ has both a very
small companion and significant vertical disc structure, hence, as was found in our modelling of the multi-colour lightcurves
of 4U1957+115 \cite{haka99}, shadowing could easily render its modulated contribution negligible.  This same shadowing could be partly
responsible for the dominant
 shorter (X-ray) period 
 optical modulation, together with both enhanced emission from and self-obscuration of the disc, all related to the negative superhump effect itself.

Moving on to X-ray dips: as in the standard formation scheme, they arise from the obscuration of the central
 X-ray source by material thrown up at the impact point of the accretion stream and the disc.  But what is novel is that the radial and azimuthal position of this region
 varies systematically with the precession of the disc. This was first put forward by \scite{scha96} to
 explain the pre-eclipse dips seen in Her X-1, which are definitively known to recur on a period less than the orbital \cite{scot99}. The
 warped disc, with a retrograde, precessing line of nodes, allows the accretion stream to overshoot the disc edge itself
 for a large fraction ({\it but not all}) of the precession cycle, leading to dips which occur progressively earlier in binary phase. 

\subsubsection{Comparison with observations}
{\it In favour:}\begin{enumerate}
\item With $P_{opt}=P_{orb}$ as was the case in the triple model, the stability of the optical modulation is consistent with its orbital origin.
\item  There is the possibility to account for the systematic dip morphology changes.  Due to the warping the accretion stream will impact the
 disc at varying angles of incidence and depths within the neutron star's potential
well (affecting the energetics).

\end{enumerate}
{\it Against:}\begin{enumerate}
\item As a corollary to the warp leading to dip shape changes, we encounter the problem that as soon as the disc leaves the mid plane, dips can only
occur for half the precession cycle when the stream intersects the face of the disc we see.  So one can explain the absence of primary dips
easily enough, but
for far too much of the cycle (as in Her X-1).

\end{enumerate}

\subsection{Summary}
Following the comparison of our observational data with a number of possible models for \targ, we find that both models invoking the optical
period as that of the binary have serious problems. In the cases of the hierarchical triple model and the permanent negative superhumper
inconsistencies are found with their predictions for X-ray dip formation.  The most convincing
model is that of a permanent positive superhumper, where a tidally distorted, elliptical disc precesses in a prograde manner, leading to the two
close periodicities, and requiring the X-ray dips to recur on the binary orbital period.  However, this promising model can not explain all
the observational features.  Specifically, more work is needed to confirm whether indeed the X-ray dip period is more stable than the optical 
(consistent with its proposed orbital origin).  This will require reanalysis of archival X-ray data, and further observational effort optically.  Lastly, this model provides no ready explanation for the dip morphology cycle and indeed none of the models
can explain the discrepancy between its 4.74\dy\ period and the 3.90\dy\ ``beat'' period of \til3000\sec\ signals.   Clearly, detailed modelling
of the combined corresponding X-ray and optical orbital modulations is needed to tackle these issues, but such work is beyond the scope of
the current paper, and will be presented in a forthcoming one.

\section{ACKNOWLEDGEMENTS}
We wish to thank the {\it RXTE} SOC team for their efforts in
scheduling the simultaneous time, and Dr Magnus N\"aslund for his kind assistance in helping us to achieve the longest possible time baseline for the NOT
observations. We are also grateful to Chris Done for providing code to generate power-law noise lightcurves, which we used to simulate the red-noise of
the ASM data.  The Nordic Optical Telescope is operated on the island of La Palma
      jointly by Denmark, Finland, Iceland, Norway, and Sweden, in the
      Spanish Observatorio del Roque de los Muchachos of the 
      Instituto de Astrofisica de Canarias.  The data presented here have been taken using ALFOSC, which is
       owned by the Instituto de Astrofisica de Andalucia (IAA) and
       operated at the Nordic Optical Telescope under agreement between
       IAA and the NBIfA of the Astronomical Observatory of Copenhagen.  This
paper also utilizes results provided by the ASM/{\it RXTE} team.
  PJH was supported by an
Academy of Finland research grant.

\bibliographystyle{../bst/mymn}

\end{document}